\documentclass[a4paper,11pt]{article}
\usepackage[T1]{fontenc}
\usepackage{jcappub}
\usepackage{graphicx}
\usepackage{dcolumn}
\usepackage{bm}

\usepackage{hyperref}
\usepackage{xcolor}
\usepackage{footnote}

\title{Non-Gaussianity from the Cross-correlation of the Astrophysical Gravitational Wave Background and the Cosmic Microwave Background}

\author[a,b]{Gabriele Perna,}
\author[a,b,c]{Angelo Ricciardone,}
\author[a,b,d]{Daniele Bertacca,}
\author[a,b,d,e]{Sabino Matarrese}

\affiliation[a]{Dipartimento di Fisica e Astronomia ``Galileo Galilei'', Universit\`a degli Studi di Padova,\\ Via Marzolo 8, I-35131, Padova, Italy}
\affiliation[b]{INFN, Sezione di Padova,\\ Via Marzolo 8, I-35131, Padova, Italy}
\affiliation[c]{Dipartimento di Fisica ``Enrico Fermi'', Universit\`a di Pisa, \\ Largo Bruno Pontecorvo 3, Pisa I-56127, Italy}
\affiliation[d]{INAF - Osservatorio Astronomico di Padova, \\ Vicolo dell'Osservatorio 5, I-35122 Padova, Italy}
\affiliation[e]{Gran Sasso Science Institute, \\Viale F. Crispi 7, I-67100 L'Aquila, Italy}

\emailAdd{gabriele.perna@phd.unipd.it}
\emailAdd{angelo.ricciardone@unipi.it}
\emailAdd{daniele.bertacca@pd.infn.it}
\emailAdd{sabino.matarrese@pd.infn.it}
\abstract{
Since the first LIGO/Virgo detection, Gravitational Waves (GWs) have been very promising as a new complementary probe to understand our Universe. One of the next challenges of GW search is the detection and characterization of the stochastic gravitational wave background (SGWB), that is expected to open a window on the very early Universe (cosmological background) and to provide us new information on astrophysical source populations (astrophysical background). One way to characterize the SGWB and to extract information about its origin is through the cross-correlation with other cosmological probes. To this aim, in this paper, we explore the cross-correlation between the astrophysical background anisotropies and the Cosmic Microwave Background (CMB) ones. Such a signal is sensitive to primordial non-Gaussianity (nG) through the GW bias. Thus, we study the capability of next generation space-based interferometers to detect such a cross-correlation signal and to constrain primordial nG.
\begin{description}
\item[Keywords] Gravitational Wave Background - Cosmic Microwave Background - non-Gaussianity 
\end{description}
}

\begin{document}
\maketitle
\flushbottom

\section{\label{Sec:Introduction}Introduction}

One of the future goals for Gravitational Wave (GW) astronomy is the detection of the Stochastic Background of Gravitational Waves \cite{Sathyaprakash:2011bh,LISA:2017pwj,LIGOScientific:2016wof,Maggiore:2019uih}.
After three runs by the LIGO-Virgo-KAGRA (LVK) collaboration we are very close to the detection of the stochastic background produced by astrophysical objects \cite{KAGRA:2021kbb,KAGRA:2021mth}. On the other hand a claim of a possible detection has been already done by the NANOGrav collaboration~\cite{KAGRA:2021kbb}, which in the 12.5 years data has found the presence of a common red-noise process \cite{NANOGrav:2020bcs}. Current operative ground-based interferometers should reach their design sensitivity in a few years and this will be probably crucial for the detection \cite{KAGRA:2021kbb}. 


Two different contributions are expected to contribute to the  background: an astrophysical component (AGWB), originated from the superposition of a large number of unresolved gravitational wave signals coming from very far or very faint astrophysical sources~\cite{Ferrari:1998jf,Ferrari:1998ut,Ignatiev:2001jr,2004MNRAS.351.1237H,Regimbau:2007ed,Regimbau:2011rp}, and a cosmological one (CGWB), coming from the very early stages of the Universe, generated by mechanisms as inflation, first-order phase transition, or scalar-induced GWs~\cite{Bartolo:2016ami,Guzzetti:2016mkm,Caprini:2018mtu, LISACosmologyWorkingGroup:2022jok}. The former will provide us information about population properties of astrophysical sources, while the latter will give us access to the early Universe's stages. Since we are close to the detection of a SGWB, it is important to understand to which extent such signal can be used to extract information about our Universe, eventually also in combination with other cosmological probes. Hence, in this paper, we try to achieve a challenging task, aiming to use the AGWB to constrain primordial nG, typically parametrized in terms of the parameter $f_{\rm NL}$ \cite{Bartolo:2004if}. More specifically we focus on primordial nG of the local type (\emph{e.g.}, see~\cite{Salopek:1990jq,Gangui:1993tt,PhysRevD.63.063002}). Typically, measurements of nG from Large-Scale Structure (LSS) have been performed by means of electromagnetic signals (\emph{e.g.}, photons emitted by galaxies), while in this work we consider a different signal, \emph{i.e.} GWs.
 
Although the isotropic component of the AGWB (\emph{i.e.} the monopole) already contains plenty of astrophysical information~\cite{Maggiore:2019uih,Bartolo:2016ami,Caprini:2019egz,Sathyaprakash:2011bh}, in our study we focus on its anisotropic contribution. Such anisotropies are imprinted both at the GW production time and during the propagation of the GW signal~\cite{Bartolo:2019oiq, Bartolo:2019yeu,Bertacca:2019fnt,Contaldi:2016koz,Cusin:2019jhg}. They share many analogies with CMB ones~\cite{dodelson2020modern,Bartolo:2006cu,Bartolo:2006fj} and they are also generated by the same common seeds (\emph{i.e.}, gravitational potentials)~\footnote{The same properties are also shared by other tracers like, for example, the Cosmic Infrared Background (CIB). In this work we focus on the CMB, since even if it is produced earlier with respect to the CIB or others tracers, its anisotropies are affected by late-Universe evolution.}. 

Since they are affected by the same large-scale (scalar and tensor) perturbations and that gravitons propagate on the same geodesics of CMB photons induces, for-free, a cross-correlation signal. Despite the two signals are generated at different epochs, such a correlation is non-negligible, as shown in~\cite{Ricciardone:2021kel}. This mainly arises from the interplay between the density contribution to the anisotropies of the astrophysical background and the Integrated Sachs-Wolfe (ISW) of the CMB \cite{1967ApJ...147...73S}. 

On the astrophysical side, we consider GWs coming from unresolved Binary Black Holes (BBHs), expected to dominate the signal and very likely to be observed by Advanced LIGO~\cite{Regimbau:2016ike}.
The quantity that links the distribution of GW sources to the underlying dark matter distribution is the bias. GW mergers, in fact, reside in galaxies, related to the distribution of halos who themselves trace the dark matter distribution. Furthermore, clusters correspond to very rare high peaks of the initial density field \cite{Bardeen:1985tr}; thus from the abundance of such objects it is possible to get information on the shape of the underlying distribution and it has been shown that the bias is affected by the presence of deviation from Gaussianity in the primordial Universe \cite{Dalal:2007cu,Matarrese:2008nc,Slosar:2008hx}.  In this work we try to extract information about nG from its effect on the GW bias. 

Primordial nG is an extremely powerful observable to understand the level of interaction of the inflaton field, and to distinguish among different inflationary models. The Planck collaboration put strong constraints on nG from CMB observations, which happens to be very small~\cite{Planck:2019kim}. Other cross-correlation signals have been used in the past to constrain nG; for instance, \cite{Tucci:2016hng} cross-correlates CIB anisotropies at different frequencies, showing that in the future it will be possible to constrain local nG down to $|f_{\rm{NL}}| < 1$; \cite{Giannantonio:2013uqa} put constraints on local nG by exploiting the cross-correlation between galaxy clustering and the ISW (see also \cite{Bellomo:2020pnw}).

In this work we focus on the large-scale effect of local primordial nG on the GW bias. The presence of local nG, in fact, induces an additional scale-dependent term in the bias scaling as $1/k^2$ and produces relevant effects on the AGWB spectrum (mainly on large scales), both on the auto- and on the cross-correlation.

To extract both the AGWB auto-correlation and the AGWB$\times$CMB cross-spectrum we modify the CLASS code \cite{2011arXiv1104.2932L}. We implement all the contributions to the AGWB anisotropies (density, Kaiser term, Doppler effect and gravitational potential contributions) and also the non-Gaussian contribution to the bias. We develop an external python module to model the astrophysical dependencies that are used to extract both the monopole AGWB energy density and the anisotropies, taking into account the latest LVK astrophysical constraints~\cite{KAGRA:2021kbb,2021arXiv211103634T}. In order to obtain the binary merger rate we start from the star formation rate (SFR) proposed by~\cite{Madau:2014bja} and we convolve it with a time-delay distribution. The modelling of the waveforms is done through the phenomenological fit proposed by~\cite{Ajith:2007kx,Ajith:2009bn,Ajith:2012mn}. Then we perform a signal-to-noise (SNR) analysis focusing on next generation space-based interferometers like LISA~\cite{LISA:2017pwj} and BBO~\cite{Corbin:2005ny}. Finally, we forecast the capability of such detectors to constrain $f_{\rm NL}$ through a Fisher matrix analysis. 

The paper is organized as follows. In Section \ref{Sec::SGWB} we introduce the formalism of the SGWB and we briefly report the steps for obtaining the anisotropies of the AGWB. Then in Section \ref{Sec::CMB} we briefly review the CMB anisotropies, fundamental part of this work. In Section \ref{Sec::Bias} we discuss about the effects of primordial non-Gaussianity on the bias and finally in Section \ref{Sec::XC} we introduce the cross-correlation formalism. In Section \ref{Sec::Discussion} we discuss our results and we conclude in Section \ref{Sec::Conclusions}.

\section{\label{Sec::SGWB}Stochastic Gravitational Wave Background}
The SGWB is usually characterized in terms of the GW energy density per logarithmic observed frequency $f_{\rm o}$, defined as
\begin{equation}
    \Omega_{\rm GW} (f_{\rm o},\hat{\mathbf{n}}) = \frac{f_{\rm o}}{\rho_{\rm c}} \frac{\rm{d} \rho_{\rm GW}}{\text{d} f_{\rm o}d\Omega_{\rm{o}}}\,,
\end{equation}
where $\Omega_{\rm o}$ is the solid angle along the line of sight $\hat{\mathbf{n}}$.
Here, $\rho_{\rm c} = 3H_0^2/(8 \pi G)$ represents the critical density of the Universe, with $H_0$ and $G$ the Hubble parameter and  the Newton's constant respectively. The total energy density includes both a background contribution, by definition homogeneous and isotropic, and a directional dependent one, which is the main focus of this work. Typically, one can also decompose the total energy density into a sum over the contributions to the GW energy density coming from different GW sources
\begin{equation}
     \Omega_{\rm GW} (f_{\rm o},\hat{\mathbf{n}}) = \sum_{\rm i}  \Omega_{\rm GW}^{[\rm{i}]} (f_{\rm o},\hat{\mathbf{n}})\,,
\end{equation}
where the index ``i'' runs over different sources that contribute to the SGWB. As anticipated, we will focus on the contribution coming from BBHs.

\subsection{\label{SubSec::GWB_Anisotr}Anisotropies}

Defining with ${\overline{\Omega}}_{\rm GW}$ the isotropic (background) energy density, the total relative fluctuation can be written as
\begin{align}
    \Delta_{\rm GW}(f_{\rm o}, \hat{\mathbf{n}}) & \equiv \frac{\Delta \Omega_{\rm GW}(f_{\rm o}, \hat{\mathbf{n}})}{\overline{\Omega}_{\rm GW}(f_{\rm o})}\nonumber\\ 
    & = \frac{\Omega_{\rm GW}(f_{\rm o}, \hat{\mathbf{n}}) -  \Omega_{\rm GW}(f_{\rm o})}{\overline{\Omega}_{\rm GW}(f_{\rm o})}\,.
\end{align}
Analogously, one can define the single contribution coming from each source, so that the total one can be obtained as a weighted sum of the single ones, with weights $w_{\rm GW}^{[\rm i]} = \overline{\Omega}_{\rm GW}^{[\rm i]}(f_{\rm o})/\overline{\Omega}_{\rm GW}(f_{\rm o})$. The anisotropic contribution was firstly derived in \cite{Cusin:2017fwz,Jenkins:2018uac,Cusin:2019jhg} and in a general gauge by \cite{Bertacca:2019fnt} applying the Cosmic Rulers formalism \cite{Schmidt:2012ne} (see also the follow up paper \cite{2022JCAP...06..030B}). In the following, we will report such a contribution in the Poisson gauge, following \cite{Bertacca:2019fnt}, since it is easier to distinguish the different terms and it is the way they are implemented in CLASS. The line element can be written as
\begin{align}
ds^2 = a^2(\eta)[-(1+2\Psi(\mathbf{x},\eta))d\eta^2+\delta_{ij}(1-2\Phi(\mathbf{x},\eta)) dx^i dx^j]\,,
\end{align}
where we indicated with $\eta$ the conformal time, $x^i$ the comoving coordinates, $a$ the scale factor and $\Psi$ and $\Phi$ the Bardeen potentials. It results~\cite{Bertacca:2019fnt,2022JCAP...06..030B,2022JCAP...11..009B}
\begin{align}
    \Delta_{\rm {GW}}^{[\rm i]}(f_{\rm{o}}, \hat{\mathbf{n}}) = & \frac{f_{\rm o}}{\rho_{\rm c}} \int \frac{dz}{H(z)} \mathcal{W}^{[\rm i]} (z)\nonumber\\
    & \quad \times\bigg\{ \delta_{\rm GW}^{[\rm i]} \nonumber\\
    & \qquad + \left(b_{\rm evo}^{[\rm i]} - 2 - \frac{\mathcal{H}^\prime}{\mathcal{H}^2} \right)\hat{\mathbf{n}}\cdot \mathbf{v} - \frac{1}{\mathcal{H}} \partial_{\parallel} (\hat{\mathbf{n}}\cdot\mathbf{v}) - (b_{\rm evo}^{[\rm i]}-3)\mathcal{H}V\nonumber\\
    & \qquad +\left( 3 - b_{\rm evo}^{[\rm i]} + \frac{\mathcal{H}^\prime}{\mathcal{H}^2} \right) \Psi + \frac{1}{\mathcal{H}}\Phi^\prime + \left( 2- b_{\rm evo}^{[\rm i]} + \frac{\mathcal{H}^\prime}{\mathcal{H}}\right) \int_0^{\chi (z)} d\chi (\Psi^\prime + \Phi^\prime)\nonumber\\
    & \left.\qquad + \left(b_{\rm evo}^{[\rm i]} - 2 - \frac{\mathcal{H}^\prime}{\mathcal{H}^2}\right) \left(\Psi_{\rm o}-\mathcal{H}_0 \int_0^{\eta_0} d\eta \frac{\Psi(\eta)}{1+z(\eta)}\bigg|_{\rm o}-\hat{\mathbf{
    n}}\cdot \mathbf{v})_{\rm o} \right)
   \right\}\,.
   \label{eq:anis}
\end{align}
Again the index ``i'' runs over different sources that contribute to the AGWB. We indicated with $\chi(z)$ the comoving distance at redshift $z$, with $\mathcal{H}$ the comoving Hubble rate. The prime indicates the conformal time derivative, while the symbol $\partial_\parallel$ a derivative along the line of sight; $\mathbf{v}$ is the peculiar velocity.
It is important to specify that the quantity $\delta^{[\rm{i}]}_{\rm{GW}}$ in equation \eqref{eq:anis} is expressed in the comoving gauge and not in the Poisson gauge, where it would not be gauge invariant. Thus, before linking it to the underlying matter density contrast as explained in the introduction, we map $\delta^{[\rm i]}_{\rm GW}$ from the Poisson to the comoving gauge (where it is gauge invariant) as
\begin{equation}
    \delta^{[\rm i](\rm{P})}_{\rm GW} = \delta^{[\rm i](\rm{SC})}_{\rm GW} - (b_{\rm{evo}}^{[\rm{i}]} -3)\mathcal{H}V\,.
\end{equation}
$b_{\rm evo}^{[\rm i]}$ is the evolution bias and accounts for the formation of new sources in time. It is defined as \cite{Jeong:2011as,Bertacca:2012tp}
\begin{equation}
    b_{\rm evo}^{[\rm i]} = \frac{\text{d} \ln \left(a^3 \frac{d N^{[\rm i]}}{\text{d}z\text{d}\Omega}\right)}{\text{d}\ln a}\,,
\end{equation}
with $N^{[\rm i]}$ the comoving number density of sources.
The function $\mathcal{W}(z)$ is the astrophysical kernel that will be described in the Appendix \ref{App::AstroKernel}. It contains all the astrophysical information regarding the sources under consideration; it acts as a weight function for each redshift considered. On the first line of Eq. \eqref{eq:anis} we report the ``density'' term. Actually it is an instrinsic anisotropy and it is due to the inhomogeneous and anisotropic distribution of GW sources in the sky. This typically gives a dominant contribution in the auto- and cross-correlation. On the second line we can distinguish the Kaiser and the Doppler terms \cite{Bertacca:2019fnt}, on the third line the gravitational potential terms and, finally, on the fourth line the terms that are evaluated at the observer (indicated with the subscript $\rm{o}$). Actually these latter give contribution only to the monopole and to the dipole, so in this analysis they have been disregarded. 

Before concluding this part we introduce the bias in our formalism. Black Hole mergers, in fact, are tracers of the underlying dark matter distribution. This translates in having a bias parameters linking their overdensity to the underlying dark matter distribution one as
\begin{equation}
    \delta_{\rm GW}^{[\rm i]} = b^{[\rm i]}_{\rm GW} \delta_{\rm m}\,.
\end{equation}
In the following we will rename $b_{\rm{GW}}^{\rm{BBH}} = b$. We conclude adding that even if in general the bias is allowed a time dependence, in this work we consider an effective bias for the BBHs fixing $b = 1.5$, following \cite{Scelfo:2018sny,2022JCAP...06..030B} (see also \emph{e.g.} \cite{Mukherjee:2019wcg,Libanore:2020fim}). 

\subsection{\label{SubSec::GWB_Angular}Angular decomposition}

In order to characterize the AGWB anisotropies we decompose them on the two dimensional sky, \emph{i.e.} on the sphere. So, we can work in harmonic space, decomposing the energy density in spherical harmonics as
\begin{equation}
    \Delta_{\rm GW} = \sum_{\ell m} a_{\ell m}^{\rm GW} Y_{\ell m} (\hat{\mathbf{n}})\,.
\end{equation}
All the physics is contained in the coefficients of the expansion $a_{\ell m}^{\rm GW}$ that are given by
\begin{align}
    a_{\ell m}^{\rm GW} & = \int d^2 \hat{\mathbf{n}} Y_{\ell m}^* (\hat{\mathbf{n}}) \Delta_{\rm GW}(\hat{\mathbf{n}})\nonumber\\
    & = \sum_{[\rm i] \alpha} \int \frac{d^3\mathbf{k}}{(2\pi)^3} Y_{\ell m}^*(\hat{\mathbf{k}}) \mathcal{S}_\ell^{[\rm i]\alpha}(\hat{\mathbf{k}})\zeta(\mathbf{k})\,.
\end{align}
The index $\alpha$ runs over the various anisotropies defined in subsection \ref{SubSec::GWB_Anisotr}, while the functions $\mathcal{S}^{[\rm i]\alpha}(\hat{\mathbf{k}})$ are called source functions; they contain the information about the anisotropies and will be an important piece in the construction of the cross-correlation.  Their full expressions can be found in~\cite{Bertacca:2019fnt}. We report as an example the one related to the density anisotropies,
\begin{align}
    \mathcal{S}_\ell^{[\rm i]\delta_m } = (4\pi) i^\ell \int \frac{dz}{H(z)}\mathcal{W}^{[\rm i]} (z) b_{\rm GW}^{[\rm  i]}  j_\ell(k\chi(z))T_{\delta_{\rm m}}(\eta,k)\,,
\end{align}
with $j_\ell(x)$ the spherical Bessel function and $T_{\delta_{\rm m}}$ is the transfer function linking $\delta_{\rm m}$ to the primordial curvature perturbation $\zeta$ and $\eta$ is the conformal time. Here $T_{\delta_{\rm m}}$ is proportional to $ D(\eta)T(k)$, where $D(\eta)$ is the growth function and $T(k)$ is the usual matter transfer function obtained with CLASS \cite{2011arXiv1104.2932L,dodelson2020modern}. The angular power spectrum is defined as
\begin{equation}
    C_\ell^{\rm AGWB} = \sum_{m = -\ell}^{\ell} \frac{\left\langle  {a_{\ell m}^{\rm GW}}^* a_{\ell m}^{\rm GW}\right\rangle}{2\ell + 1} = \sum_{i,j;\alpha\beta}C_\ell^{\rm [i,j]\alpha\beta}\,.
\end{equation} 
It is straightforward to show that
\begin{equation}
    C_\ell^{\rm [i,j]\alpha\beta} = \int \frac{dk}{(2\pi)^3} k^2 {\mathcal{S}_\ell^{[\rm i] \alpha}}^* (k)\mathcal{S}_\ell^{[\rm j] \beta}(k) P_\zeta(k)\,,
\end{equation}
where $P_\zeta(k)$ is the primordial power spectrum (\emph{i.e.} the two-point correlation function in Fourier space). The dimensionless power spectrum can be built from $P_\zeta(k)$ as \cite{Planck:2018vyg,dodelson2020modern}
\begin{align}
    \mathcal{P}_\zeta(k) = \frac{k^3}{(2\pi^2)} P_\zeta(k) = \mathcal{A}_{\rm s} \left(\frac{k}{k_{\rm p}}\right)^{n_{\rm s}-1}\,,
\end{align}
where $A_{\rm s}$ and $n_{\rm s}$ are the scalar perturbations amplitude and tilt, respectively, and $k_{\rm p}$ is the pivot scale.

\section{Cosmic Microwave Background}
\label{Sec::CMB}
 The Cosmic Microwave Background has been found in 1964 by Penzias and Wilson, that were able to see just the monopole~\cite{Penzias:1965wn}. Up to now we have been able to characterize both the monopole and the anisotropies of the signal with very high accuracy (up to $\ell \sim 2500$)~\cite{Planck:2019nip}.  Since the description and characterization of the CMB is out of the purposes of this work, but at the same time is an important part of the work, we will discuss just the main points, reporting the reader to specific works for a more detailed analysis. We start from defining the CMB temperature as~\cite{1995PhDT..........H}
 \begin{equation}
     T(\mathbf{x},\hat{\mathbf{n}},\eta_0) = T(\eta_0) [1+\Theta(\mathbf{x},\hat{\mathbf{n}},\eta_0)]\,.
 \end{equation}
Here the anisotropies are contained in the function $\Theta$. The three main contributions to the CMB anisotropies on large scales are the Sachs-Wolfe, the Doppler and the Integrated Sachs-Wolfe (ISW). Actually the first one is imprinted at the moment of the decoupling and is a gravitational redshift effect. The Doppler depends on the peculiar velocity of the Earth with respect to the cosmic fluid, while the third one is still a gravitational redshift effect, but imprinted during the propagation of the signal (we will discuss this latter effect in more detail later). Also for the CMB the typical procedure is to expand the perturbation in spherical harmonics as
\begin{equation}
    \Theta(\mathbf{x}_{\rm o},\hat{\mathbf{n}},\eta_0) = \sum_{\ell m} Y_{\ell m}^{*}(\hat{\mathbf{n}}) a_{\ell m}^{\rm CMB}\,,
\end{equation}
with $\mathbf{x}_{\rm o}$ position of the observer.
At the end one gets
\begin{align}
     a_{\ell m}^{\rm CMB} & = \int d^2 \hat{\mathbf{n}} Y_{\ell m}^* (\hat{\mathbf{n}}) \Theta(\hat{\mathbf{n}})\nonumber\\
     & = \int \frac{d^3\mathbf{k}}{(2\pi)^3}Y_{\ell m}^*(\hat{\mathbf{k}}) \mathcal{T}_\ell(\hat{\mathbf{k}})\zeta(\mathbf{k})\,.
\end{align}
Here we indicated with $\mathcal{T}_\ell(\hat{\mathbf{k}})$ the source functions and also in this case there will be one for each term of the anisotropies. We report as an example the source function for the ISW, expected to dominate the cross-correlation with the density anisotropies described before, \emph{i.e.}
\begin{equation}
    \mathcal{T}_\ell^{\rm ISW} = (4\pi) i^\ell \int_0^{\eta_0} d\eta \left[ T_{\Psi}^\prime(\eta,k) + T_{\Phi}^\prime(\eta,k) \right]e^{-\tau} j_\ell(k(\eta_0-\eta))\,.
\end{equation}
In this equation $T_{\Phi}$ and $T_{\Psi}$ are transfer functions that relate the gravitational potential to the primordial curvature perturbation $\zeta$ (for the explicit expression see \emph{e.g.} \cite{dodelson2020modern}). The $j_\ell$'s are the spherical Bessel functions, $\eta$ the conformal time and $\tau$ the optical depth.

\section{\label{Sec::Bias}Non-Gaussian Bias}

The quantity that links the distribution of GW sources to the underlying dark matter distribution is the bias. In this work we focus on the effects of primordial local nG on it. In this case the primordial gravitational potential can be modeled as \cite{Gangui:1993tt,PhysRevD.63.063002}
\begin{equation}
\label{Eq::NGPotential}
    \Phi = \Phi_{\rm L} + f_{\rm NL}(\Phi_{\rm L}^2 - \langle \Phi_{\rm L}^2 \rangle)\,,
\end{equation}
where $\Phi_{\rm L}$ is the Gaussian potential, while $f_{\rm NL}$ is a parameter that quantifies the amount of primordial nG. Actually, starting from \eqref{Eq::NGPotential} one can show that the effects of nG can be recasted in an additional scale-dependent contribution to the bias $\Delta b$ such that \cite{Slosar:2008hx, Matarrese:2008nc,Dalal:2007cu}
\begin{equation}
\label{Eq::NG_correction}
    \Delta b(k) = 3 f_{\rm NL} (b-1) \delta_{\rm c} \frac{\Omega_{\rm m}}{T(k) D(z)k^2} \left( \frac{H_0}{c}\right)^2\,.
\end{equation}
Here $D(z)$ is the growth function, $T(k)$ the matter transfer function and $\delta_c = 1.686$ is the spherical collapse linear over-density \cite{Kaiser:1984sw,Bardeen:1985tr,Efstathiou:1988tk}.
Such a correction is proportional to $f_{\rm NL}$, meaning that in principle we could use this signal to constrain this parameter. Moreover we observe that the scale dependence goes like $k^{-2}$ meaning that the effect of the correction will be important on large scales (small $k$). This is an important remark since actually we expect our cross-correlation signal to be sizeable mainly on large-scales and so sensitive to this correction. This strengthens the idea that we could use this signal to constrain nG.

\begin{figure}[t!]
\centering
\includegraphics[width=0.8\linewidth]{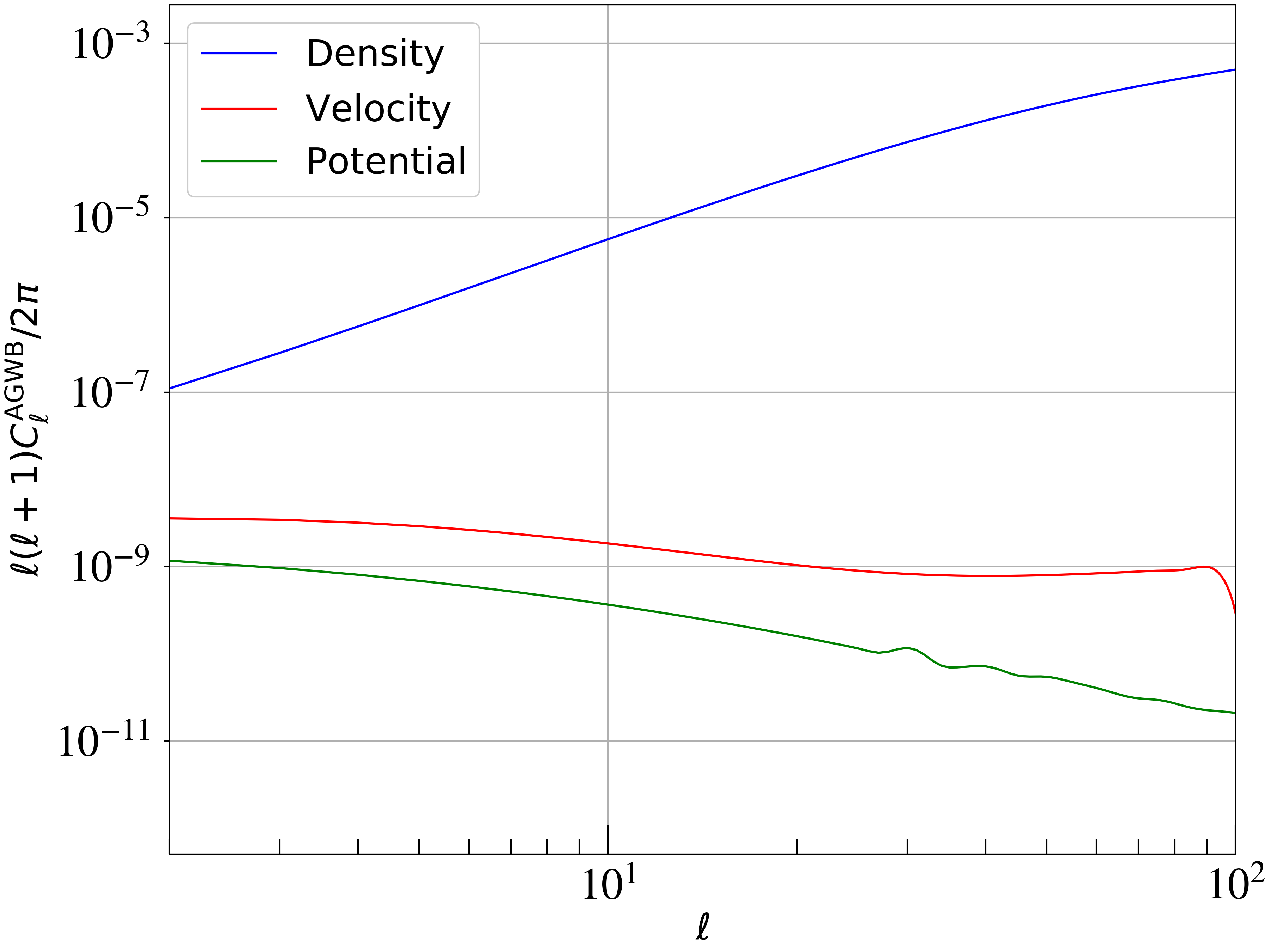}  
\caption{\label{Fig::AGWB-Auto} The plot shows the auto-correlation angular power spectrum of the AGWB anisotropies. As expected, the density contribution (blue line) dominates the spectrum, even if the velocity terms (i.e., Kaiser and Doppler) provide a non-negligible contribution, especially at the large scales. The green line represents the gravitational potential contribution. We recall that the plot has been obtained modifying the public code CLASS \cite{2011arXiv1104.2932L} considering only the BBH contribution.}
\end{figure}

Actually, before proceeding, it is interesting to underline that the authors of \cite{Slosar:2008hx} specify how equation \eqref{Eq::NG_correction} is not always valid. Summarizing, one can re-write the bias as
\begin{align}
\label{Eq::General_NG_Bias}
    b_{\rm NG}\delta & = b\, \delta + b_\phi f_{NL} \phi \nonumber\\
    & = b\, \delta + b_\phi f_{NL} \frac{\delta}{\alpha}\,,
\end{align}
with $\alpha\propto T(k)D(z)$ and $b_\phi = 2\delta_c(b-1)$. This latter is called universality relation, but it is not always true. One can show that a more general relation would be $b_\phi = 2\delta_c(b-p)$, with $p$ depending on the halo properties, \emph{e.g.} if it is the result of a recent merger, if it is an old halo etc. In our case we are accounting for astrophysical sources of GW. Being formed typically at the end of a star life, they are located in galaxies with a very intense and recent stellar activity. As argued by \cite{Slosar:2008hx} (see also \emph{e.g.}, \cite{Barreira:2021ueb,Lazeyras:2022koc}), finding the right value for $p$ would make a difference in constraining $f_{\rm NL}$. In order to understand this better, we evaluate $\sigma_{f_{\rm NL}}/f_{\rm NL}$ for different values of $p$, \emph{i.e} we consider the two extreme cases $p = 1$ and $p = 1.6$ in~\cite{Slosar:2008hx}, since they are obtained through analytical estimation (rather than simulations) and are independent of the halo mass, to check how the constraints on $f_{\rm NL}$ would vary. In general the proper value of $p$ depends on the link between the properties of the tracer considered and the properties of the host halo.
We further specify that in our analysis we fix the value of $p$ instead of choosing a prior and then marginalize over it, due to the degeneracy between the $p$ and $f_{\rm NL}$. Moreover, marginalizing over $p$ is quite challenging, because its (but more generally $b_\phi$) degeneracy with $f_{\rm NL}$ makes it difficult to constrain both of them simultaneously. On the other hand, also the marginalization is not trivial. As shown in \cite{Barreira:2022sey}, for example, the choice of the prior even if quite wide, could bias the final result for $f_{\rm NL}$. Following \cite{2022JCAP...06..030B} we work with time- and host- independent PDFs for the astrophysical parameters, since interested in giving an order of magnitude estimate of the signal. 
We perform the analysis only with BBO, since the SNR resulting from LISA is too low to perform a Fisher analysis, as we show in Appendix \ref{App::SNR}.
We conclude underlying that we perform our analysis in a full GR context, in agreement with~\cite{Jeong:2011as}.

\begin{figure}[t!]
\centering
\includegraphics[width=0.8\linewidth]{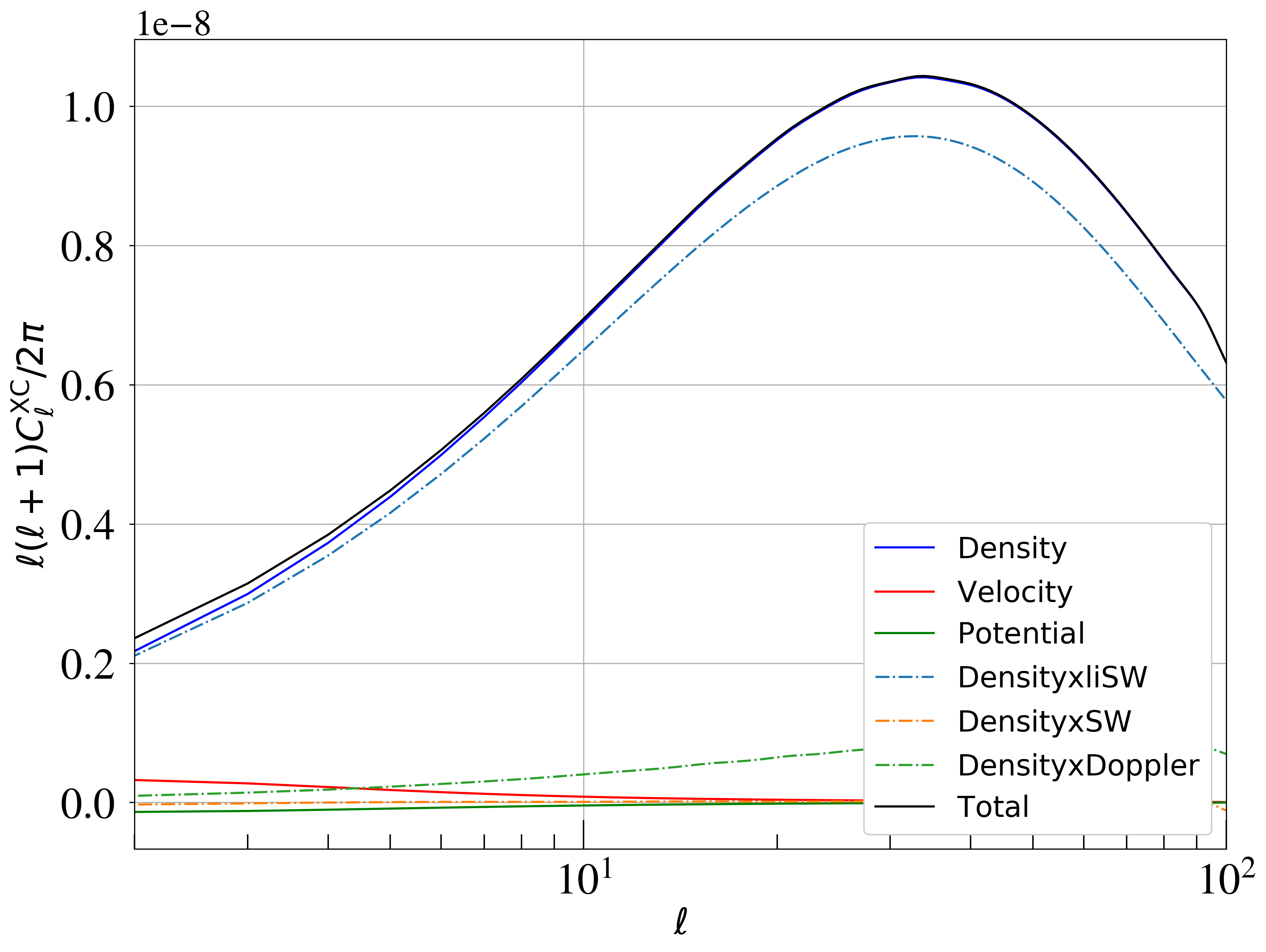}  
\caption{\label{Fig::AGWB-Cross} The plot shows the cross-correlation angular power spectrum of the AGWB with the CMB. The dominant contribution, as expected, comes from the CMB ISW with the density anisotropies of the AGWB (light blue dashed line). The black line represents the total contribution. We note that on large scales the velocity anisotropies provide a non-negligible contribution in the signal, while on smaller ones dominates the correlation with the density ones.}
\end{figure}

\section{\label{Sec::XC}Cross-Correlation formalism}

We evaluate the cross-correlation of the AGWB with the CMB, including the effect of nG in the bias. Actually the cross-correlation arises mainly from the anisotropy related to the in-homogeneous distribution of GW sources in the sky and the late-ISW effect for the CMB. Let us better understand the physics behind this: BBH are relatively recent sources in the Universe evolution and as remarked before they trace the underlying dark matter distribution; so these sources reside in the large-scale gravitational potential. On the other hand, the late-ISW effect is first of all very recent (it is non-vanishing since the cosmological constant started to dominate the energy density) and also dominant on large scales. It is due to the time-variation of the gravitational potentials, that are constant during matter domination, but start varying again during Dark Energy domination (and this explains why it is recent). Moreover, as the perturbation re-enters the horizon it is damped by the expansion of the Universe and this justifies why the larger the scale, the higher the amplitude of the perturbation is expected.  Let us now think for a moment on the effects we expect when adding the nG correction to the bias (equation \eqref{Eq::NG_correction}) in the analysis. Assuming to be in the case in which $(b-p)>0$, the effects of nG are expected just to increase (positive $f_{\rm NL}$) or suppress (negative $f_{\rm NL}$) the cross-correlation spectrum, while in the opposite case $(b-p)<0$ the effects are just inverted (in the range of $f_{\rm NL}$ allowed by Planck). Physically, this is due to a variation of the clustering properties of the tracers considered induced by primordial nG itself and it is expected to have an impact on the cross-correlation, but also on the constraints on $f_{\rm NL}$.

\subsection{Prescription}
In order to extract the cross-correlation signal we start by considering the single $a_{\ell m}$. For the AGWB we have 
\begin{equation}
    a_{\ell m}^{\rm GW} = s_{\ell m}^{\rm GW} + n_{\ell m}^{\rm GW}\,, 
\end{equation}
where $s_{\ell m}$ is the contribution coming from the actual signal, while $n_{\ell m}$ refers to the noise of the detector. In the case of the CMB we have neglected the instrumental noise of the CMB experiment, since at the multipoles of interest $(\ell \leq 100)$, available CMB data are completely cosmic variance dominated, \emph{i.e.} $a_{\ell m}^{\rm CMB} = s_{\ell m}^{\rm CMB}$. 

The angular cross-correlation spectrum results
\begin{align}
\label{Eq::CrossCl}
    \langle s_{\ell m}^{\rm GW} s_{\ell m}^{\rm CMB} \rangle = \delta_{\ell \ell'} \delta_{m m'} C_{\ell}^{\rm XC}\,.
\end{align}
Note that the term $\langle n_{\ell m}^{\rm GW} s_{\ell m}^{\rm CMB} \rangle$, vanishes since the noise of the instrument and the CMB signal are uncorrelated. In general the $C_\ell$ coming from  \eqref{Eq::CrossCl} are given by (notice the presence of the source functions defined before)
\begin{align}
    C_\ell^{\rm XC} = \sum_{[\rm i]\alpha} \int \frac{dk}{(2\pi)^3}k^2 \mathcal{S}_\ell^{[\rm i]\alpha}(\hat{\mathbf{k}})\mathcal{T}_\ell^*(\hat{\mathbf{k}}) P_\zeta(k)\,.
\end{align}
Of course one should write the $C_\ell$ as a sum of all the possible combinations among the various contributions seen before. Here we report the explicit expression for the dominant one, \emph{i.e.}\footnote{Recall that we are considering only the case of BBH, with an effective bias $b = 1.5$.}
\begin{align}
\label{Eq::Cl_DxISW}
    C_\ell^{\rm ISW x \delta_{\rm m}} & = \frac{2}{\pi} \sum_{[\rm i]} \int dk k^2 P_\zeta(k) \int_0^{\eta_0} d\eta \left[T_{\Psi}'(\eta,k) + T_{\Phi}'(\eta,k)\right] e^{-\tau} j_\ell(k(\eta_0 - \eta))\nonumber\\
    & \quad\times \int_{0}^{\eta_0} d\eta \mathcal{W}^{[\rm i]}(\eta) \left[ b + 3 f_{\rm NL}(b-p)\delta_{\rm c} \frac{\Omega_{\rm m}}{T(k)D(\eta)k^2}\left(\frac{H_0}{c}\right)^2 \right] j_\ell(k(\eta_0-\eta))T_{\delta_{\rm m}}(\eta,k)\,.
\end{align}
On the second line, the integral runs from 0 to today. Actually the astrophysical kernel acts as a window function that cuts the integral to recent times (star formation has not been active since the beginning of the Universe). Thus the lower bound can be simply replaced with some $\eta_{\rm min}$, depending on the model chosen for the star formation. $T_{\delta_{\rm m}}$ is the transfer function for the density contrast, $T_\Psi$ and $T_\Phi$ are the transfer functions for the gravitational potentials defined before.

\subsection{Numerical Implementation}
To obtain the cross-correlation angular spectra one should solve the Boltzmann equations for the perturbations from inflation up to today and then evaluate the various integrals as shown in \eqref{Eq::Cl_DxISW}. For this purpose we modify the publicly available code CLASS \cite{2011arXiv1104.2932L,Blas:2011rf,DiDio:2013bqa} where the CMB anisotropies are already implemented. We add the AGWB ones including the presence of the primordial nG correction to the bias. We also implement the possibility to use a time-dependent bias  as an input and the possibility to vary the $p$ parameter in the nG additional contribution. We fix the bias to 1.5 and we allow for two values of $p$ ($p = 1$ and $p = 1.6$).

Moreover we develop a python module to model and evaluate the astrophysical dependencies. Actually such kernel is built with different moduli, in such a way it can be easily adapted to include other GW sources and astrophysical dependencies. More details on the astrophysical part are given in Appendix \ref{App::AstroKernel}.

\begin{figure}[t!]
\centering
\includegraphics[width=0.8\linewidth]{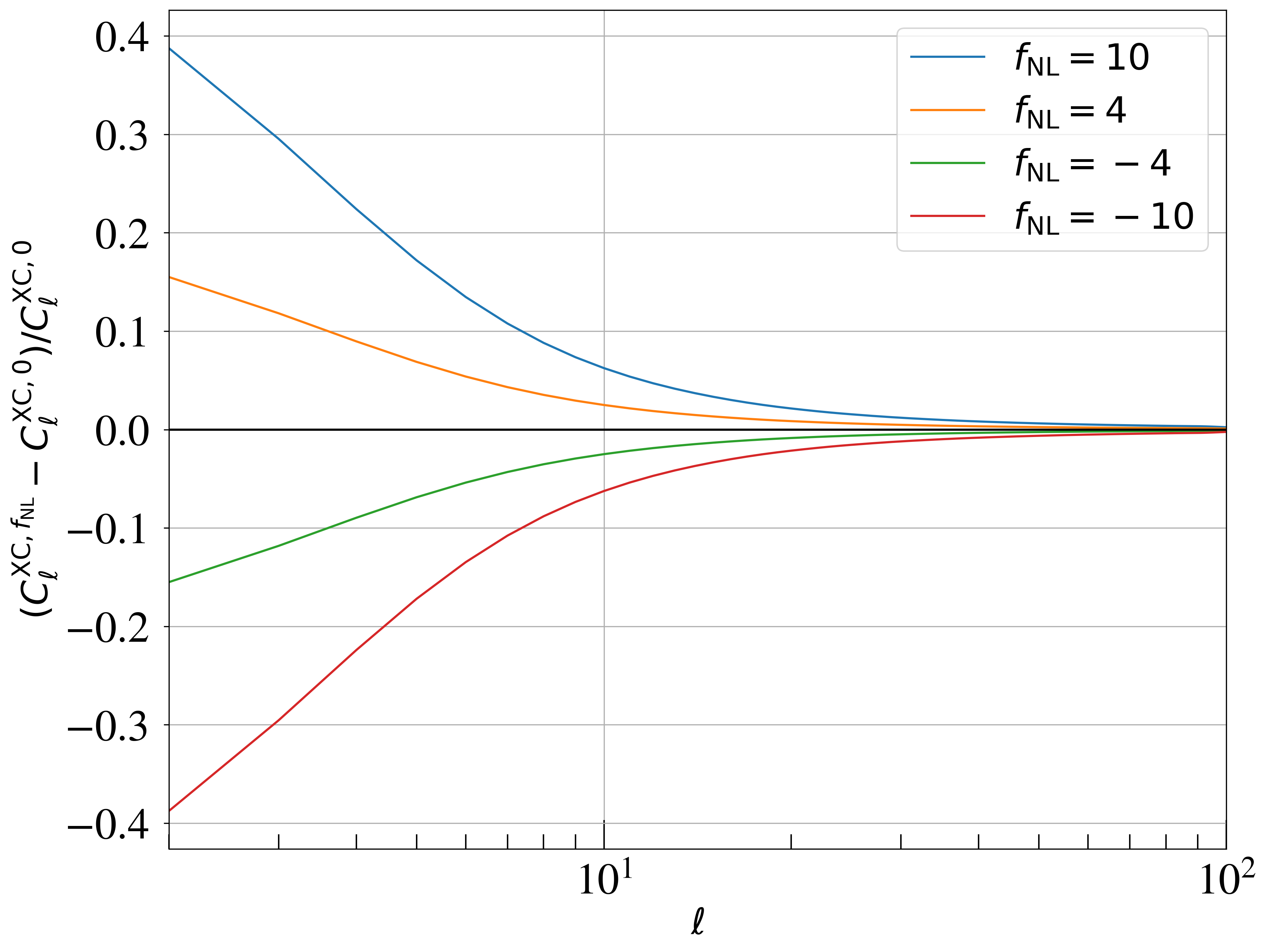}  
\caption{\label{Fig::Percentage_Cross} The plot shows the percent difference of the angular power spectrum for the cross-correlation between the AGWB and the CMB in the nG cases with respect to the Gaussian one. Actually we consider different values of $f_{\rm NL}$, showing that the nG contribution leads to a suppression (or enhancement) up to $40\%$ on large scales. The exact behaviour of the curves is actually dependent on the clustering properties of the GW sources, \emph{i.e.} on the parameters $b$ and $p$ under consideration. In this case we consider $b = 1.5$ and $p = 1$.}
\end{figure}

\section{Results\label{Sec::Discussion}}
We describe in the following subsections the plots and the results we obtained in the evaluation of the cross-correlation spectra and in the constraining of nG.

\subsection{Auto- and Cross- spectra}

We plot in Fig. \ref{Fig::AGWB-Auto} and Fig. \ref{Fig::AGWB-Cross} the AGWB auto-correlation and the AGWBxCMB cross-correlation angular power spectra. We plot the spectra up to $\ell\sim 100$, since it is unlikely that next generation GW detectors will be able to reach smaller scales. In the auto-correlation we observe how the density term dominates the spectrum, even if the velocity terms (Kaiser and Doppler) are expected to give a valuable correction \cite{Bertacca:2019fnt}. We specify that this behaviour is of course model dependent; more specifically the information about the source population comes mainly in the bias parameter, but also in the evolution bias described before. Furthermore, the behaviour of the auto-correlation and in particular of the density term is useful for a qualitative prediction of the cross-correlation spectrum, as will be explained below. 
Let us now focus on the cross-correlation spectrum. As anticipated before, we expect that the main contribution to such a spectra comes from the ISW-density term. The ISW, in particular its late contribution, results to be higher at the largest scales, to then decrease at smaller ones. On the other hand, the density anisotropy grows as the scales decrease. So it is clear that a cross-correlation among the two signals is expected to grow at the lowest $\ell$s (following the density behaviour) and then, after reaching a maximum, decrease following the dominating ISW behaviour. Fig. \ref{Fig::AGWB-Cross} shows the result of the numerical computation performed with the modified version of CLASS that confirms the expectations. The cumulative effects of the density cross-correlations, \emph{i.e.} the correlation of all the CMB terms with the density anisotropies described before (blue line), clearly dominates the signal. The total contribution (black line) and the density almost coincide on all the scales, even if we observe that on the largest scales the correlation with the velocity anisotropies (red line) is non-negligible, giving a visible contribution (few percent) on the total signal. Among the density anisotropies, indicated with dash-dotted lines in the Figure, we observe that actually the density-ISW one prevails (cyan line), almost coinciding with the total density term (blue line), especially on large scales.

We specify that in principle the contribution coming from the cross-correlation of the density anisotropies with the Sachs-Wolfe could be important on scales $k\ll \mathcal{H}/c$ and should be accounted \cite{Desjacques:2020zue}. In this analysis we integrate starting at $k\sim 0.1 \,\mathcal{H}/c$ (see also \cite{Bertacca:2012tp}) and for low values of $f_{\rm NL}$ such a contribution can be neglected, at first approximation. We leave a further detailed analysis of this effect to a future work.

In Figure \ref{Fig::Percentage_Cross} we fix the bias at a reference value $b = 1.5$ and $p = 1$, and we plot the percent difference of the cross-correlation spectrum with respect to the Gaussian case due to different values of $f_{\rm NL}$.  For positive values the spectrum increases, up to 40\% in the case in which $f_{\rm NL} = 10$, while for the negative ones we observe a suppression of the spectrum on the largest scales of the same amount\footnote{We specify that we consider $f_{\rm NL} = 10,-10$, mainly for illustrative purposes, since Planck collaboration has been able to constrain this parameter to lower values (in absolute value) \cite{Planck:2019kim}. }. This symmetry in the behaviour of the spectrum when varying from positive to negative values of $f_{\rm NL}$ is due to the peculiar linear dependence of the cross-correlation on such a parameter. The auto-correlation, for example, will present both a quadratic term in $f_{\rm NL}$ $\propto \left(f_{\rm NL}(b-p)\right)^2$ that will always give a positive contribution and a linear one that goes like $f_{\rm NL}(b-p)$. This latter on the other hand will behave as in the cross-correlation case. It is clear that the combination of the two leads to a break of the symmetry of the nG effects on the auto-correlation. 

Finally we specify that the effects on the spectrum depend also on the values chosen for $b$ and $p$. At fixed $p$, if $b>p$ and $f_{\rm NL}$ is positive we get the behaviour described above, but for a bias $b<p$, actually the effects on the power spectrum are reversed. We stress that what really matters is the whole $f_{\rm NL}(b-p)$ contribution and not only $f_{\rm NL}$: in the case in which $b \simeq p$, it would not be possible to appreciate the effects of nG. This further underlines the importance of a precise modelling of the bias. Moreover, accounting for a time-dependence in the bias could make the difference, since in principle it could be greater than $p$ at a certain time and less than $p$ in another, leading to interesting effects in the behaviour of the power spectra. This of course would need to take into account the evolution of the GW sources but also of the halos in which they reside. We leave such a detailed analysis to a future work. 

\subsection{Fisher Forecast}
We perform a Fisher forecast to provide estimates of the capability of future GW detectors to constrain $f_{\rm NL}$ through the AGWBxCMB cross-correlation. The Fisher matrix, in fact, provides a powerful tool for describing the information content coming from observables in terms of underlying parameters; it can be written as~\cite{Tegmark:1999ke}
\begin{align}
    \label{Eq::Fisher}
    F_{\alpha\beta} & = \sum_\ell^{\ell_{\rm max}} \frac{2\ell+1}{2} \mathrm{Tr}\left[\frac{\partial {C}_\ell}{\partial\theta_\alpha} {C}_\ell^{-1}\frac{\partial {C}_\ell}{\partial\theta_\beta} {C}_\ell^{-1}\right]\nonumber\\
    & = \sum_\ell^{\ell_{\rm max}} \frac{\partial \mathbf{C}_\ell^T}{\partial\theta_\alpha}\mathbb{C}_\ell^{-1}\frac{\partial \mathbf{C}_\ell}{\partial\theta_\beta}\,,
\end{align}
where in the last step we have defined $\mathbf{C}_\ell^T = (C_\ell^{\rm CMB}, C_\ell^{\rm AGWB}, C_\ell^{\rm AGWB\times CMB})$, $\mathbb{C}_\ell$ being the covariance matrix and $\theta_\alpha$, $\theta_\beta$ the parameters. In our case we have just one single parameter, \emph{i.e.} $f_{\rm NL}$. Since the $C_\ell^{\rm CMB}$ do not depend on the parameter we are willing to estimate, equation \eqref{Eq::Fisher} reduces to
\begin{equation}
\label{Eq::Fisher_Final}
F = \sum_\ell^{\ell_{\rm max}}\left(\,0\,,\frac{\partial C_\ell^{\rm AGWB}}{\partial f_{NL}}\,,\frac{\partial C_\ell^{\rm XC}}{\partial f_{NL}}\right)\mathbb{C}_\ell^{-1}
\begin{pmatrix}
0\\
\frac{\partial C_\ell^{\rm AGWB}}{\partial f_{NL}}\\
\frac{\partial C_\ell^{\rm XC}}{\partial f_{NL}}
\end{pmatrix}\,.    
\end{equation}

\begin{figure}[t!]
\centering
\includegraphics[width=0.8\linewidth]{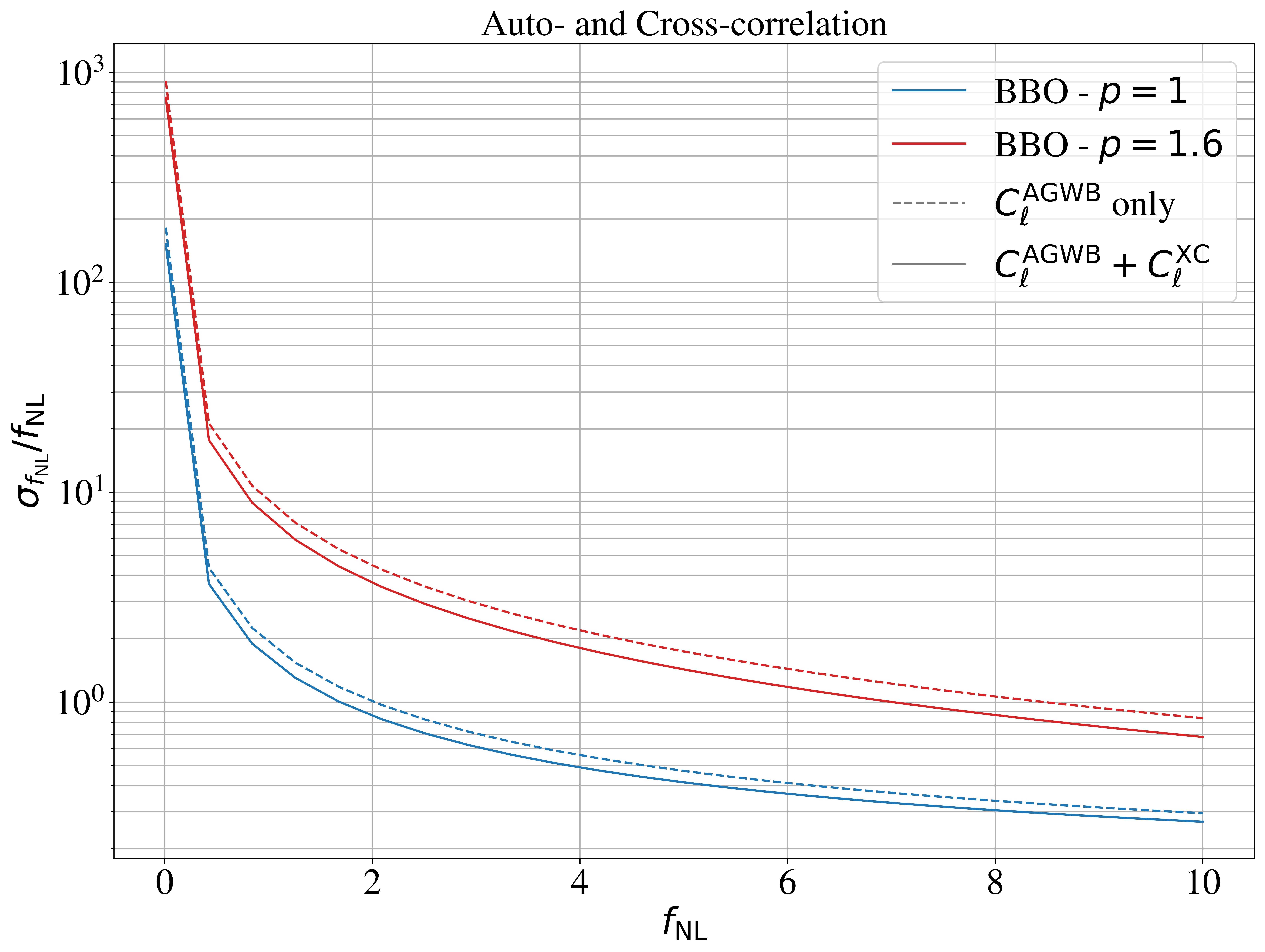}  
\caption{\label{Fig::sigma_f_NL} The plot shows the variation of $\sigma_{f_{\rm NL}}/f_{\rm NL}$ as function of $f_{\rm NL}$ for the BBO detector for different values of $p$ (blue lines for $p = 1.6$ and red lines for $p = 1$). Dashed lines refer to the inclusion in the analysis of only the $C_{\ell}^{\rm AGWB}$, while solid ones show the case in which also the $C_{\ell}^{\rm XC}$ are added. In this latter case it is possible to appreciate the non-negligible improvement in the value of $\sigma_{f_{\rm NL}}$. 
Since the exact value of $p$ is strongly dependent on the evolution properties of the halos and of the GW sources tracing them, we show how the variation of this parameter impacts on the constraining power of the signal.}
\end{figure}

We have included the angular spectrum of the instrumental noise $N_\ell$ of the interferometers in the covariance matrix. Finally one obtains that the error is given by
\begin{align}
    \sigma_{f_{\rm NL}} \simeq \sqrt{F^{-1}}\,.
\end{align}
In Fig. \ref{Fig::sigma_f_NL} we plot the $\sigma_{f_{\rm NL}}/f_{\rm NL}$ as function of $f_{\rm NL}$ for two different values of $p$, \emph{i.e.} $p=1$ (blue lines) and $p=1.6$ (red lines). We keep fixed the bias to the reference value of 1.5 and we consider the BBO detector in the evaluation of the covariance for each multipole. Furthermore we show the case in which only the $C_{\ell}^{\rm AGWB}$ are considered in the analysis (dashed lines) and the one in which also the $C_{\ell}^{\rm XC}$ are included, to show the impact of these latter on the value of $\sigma_{f_{\rm NL}}$.
Before discussing the effects of different values of $p$, we highlight that we get a higher constraining power for higher values of $f_{\rm {NL}}$. The low angular sensitivity of LISA leads to a low SNR, as shown in Appendix \ref{App::SNR}, suggesting, in this case, a poor constraining power. On the other hand, as we can see in Fig. \ref{Fig::sigma_f_NL},  BBO will be able to put tight constraints on the $f_{\rm NL}$ parameter, being able to reach percent error levels.

Let us now discuss the impact of different values of $p$ on $\sigma_{f_{\rm NL}}$. We can see that in the case $p = 1$ we are able to put tighter constraints with respect to $p = 1.6$ case. This behaviour is of course dependent on the value of the bias and on the parameters chosen for modelling the clustering properties of the tracers. In this case, fixing the bias $b$ to 1.5, it is clear that for $p = 1.6$ the effect of nG are suppressed while for $p = 1$ they are enhanced and so in this latter case we achieve a better constraining power. Thus, the plot shows how a good modelling of the sources is crucial in constraining  $f_{\rm NL}$. Furthermore, the difference between the dashed and the solid lines in the plot shows the importance of the cross-correlation. The inclusion of $C_{\ell}^{\rm XC}$ leads to a significative improvement in the value of $\sigma_{f_{\rm NL}}$. We can conclude, then, that not only the cross-correlation among the AGWB and the CMB is non-negligible, but it is also relevant for the $f_{\rm NL}$ constraints.

\begin{figure}[t!]
 \centering
\includegraphics[width=0.8\linewidth]{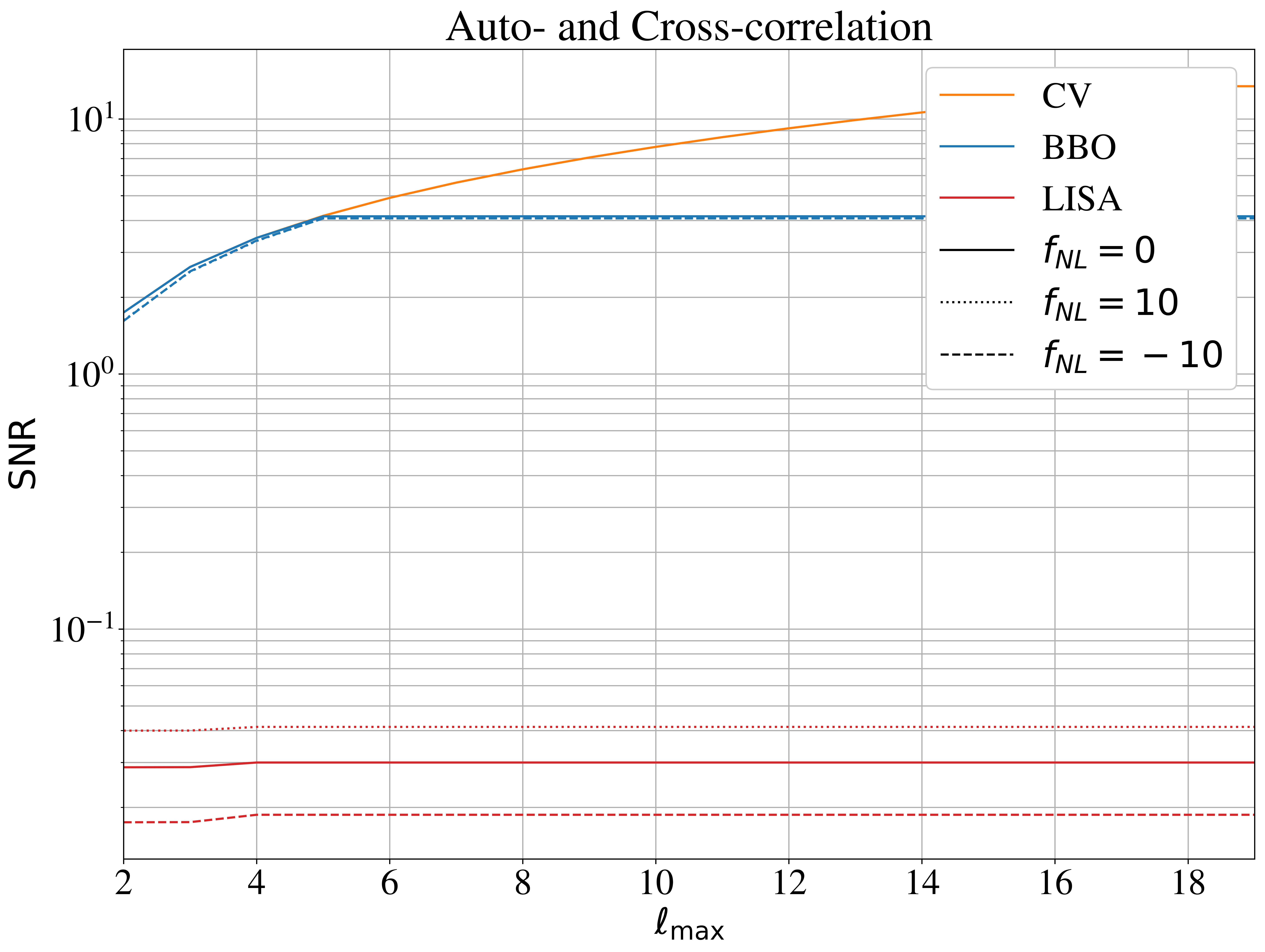}  
\caption{\label{Fig::SNR_ellmax} The plot shows the cumulative SNR obtained for different values of $\ell_{\rm max}$ in the cosmic variance-limited case (orange line), accounting for the BBO noise (blue lines) and for the LISA one (red lines). We also account for the effect of different values of $f_{\rm NL}$ on the SNR, showing that they potentially enhance or suppress it.}
\end{figure}

Then we compute the SNR for both LISA and BBO (see Appendix \ref{App::SNR} for more details on the SNR analysis). In Figure \ref{Fig::SNR_ellmax} we plot the SNR as a function of $\ell_{\rm max}$ for LISA and BBO. We can see that we are able to gain information mainly from the very first multipoles (up to $\ell \sim 4-5$),  meaning that, in principle, we could achieve a greater constraining power on $f_{\rm NL}$, if a better angular resolution could be reached.  
 We have included only the terms coming from the auto- and cross- correlation parts (\emph{i.e.}  $C_\ell^{\rm AGWB}$ and $C_\ell^{\rm XC}$). 
We can see that LISA will not be able to reach SNR larger than one, even for large (and positive) values of $f_{\rm NL}$. On the other hand, in the case of BBO, we can see that we are able to reach SNR larger than one for 5 years of observation. This reflects on  the ability of constraining $f_{\rm NL}$. Furthermore, we stress that this behaviour is strongly dependent on the choices made for both the cosmological and the astrophysical parameters, underlying that a better modeling is necessary.

 \section{Conclusions \label{Sec::Conclusions}}

In this paper we analysed the effects of local primordial nG on the cross-correlation between  CMB and AGWB anisotropies. We explored to which extent it is possible to use astrophysical GW sources to constrain early-universe cosmology. 
We introduced the theoretical formalism for describing the CMB and AGWB anisotropies and then we discussed the cross-correlation signal. This is expected to be dominated by the correlation between the density anisotropies of the AGWB and the CMB (late)-ISW effect, both effects being very recent and important on large scales. We confirmed such a behaviour by plotting the expected angular power spectrum obtained modifying the publicly available code CLASS. We added the effects of local primordial nG on the bias and the AGWB anisotropies and we modelled the population of BBHs accounting for the latest LVK astrophysical constraints. We found that the inclusion of primordial nG would lead to an enhancement (or a suppression) of the power spectrum, mainly on large scales.
Then we performed a Fisher forecast on the $f_{\rm NL}$ parameter to understand to which extent future space-based interferometers (\emph{e.g.} and BBO) will be able to put constraints on nG. We find that BBO will be able to put very tight constraints on $f_{\rm NL}$. This behaviour is confirmed by the SNR analysis, which clearly shows that BBO can reach a SNR $\sim$ 4 for 5 years of observation. On the other hand the SNR for LISA results quite low.

Our analysis shows the importance of the clustering properties of GW sources and their impact on the signal. We considered the simplified case of a constant bias and $b_\phi \propto (b-p)$, obtaining that positive values of $f_{\rm NL}$ enhance the spectrum, while negative values suppress it. We stress that this is not a general behaviour, meaning that the enhancement/suppression depends not only on the value of $f_{\rm NL}$, but also on the clustering properties of the sources. An accurate modelling of such quantities is crucial for constraining $f_{\rm NL}$. A time-dependent bias and a better modelling of the $p$ parameter would impact our analysis, avoiding under- or over-estimation of the error bars. \footnote{Let us also mention that a cross-correlation signal is expected also for resolved GW sources. However, in this paper we focused on the AGWB mainly for two reasons: the stochastic background is expected to extend up to high redshift, while resolved sources,  are expected to be closer. On the other hand, the large uncertainity on the modelling of resolved sources could have a large impact on the analysis.} 

We conclude stressing that, even if GW detectors are limited in angular sensitivity to the first few multipoles, they can play an important role in constraining early-universe cosmology.

\acknowledgments
We thank Lorenzo Valbusa Dall'Armi for useful discussions and feedback. A.~R. acknowledges financial support from the Supporting TAlent in ReSearch@University of Padova (STARS@UNIPD) for the project “Constraining Cosmology and Astrophysics with Gravitational Waves, Cosmic Microwave Background and Large-Scale Structure cross-correlations’'. D.B. and S.M. acknowledge partial financial support by ASI Grant No. 2016-24-H.0.

\appendix

\section{SNR Evaluation\label{App::SNR}}
In the main text we reported the results for the computation of the expected AGWB$\times$CMB signal at different values of $f_{\rm NL}$. We focus in this Appendix on its detectability by means of a SNR analysis. We evaluate the SNR accounting for the contributions coming from both the auto- and the cross-correlations, \emph{i.e.} 
\begin{equation}
\label{Fig::SNR}
    SNR^2  = \sum_{\ell = 2}^{\ell_{\rm max}} (C_\ell^{\rm CMB},C_\ell^{\rm AGWB}, C_\ell^{\rm XC})  \mathbb{C}_\ell^{-1}
     \begin{pmatrix}
    C_\ell^{\rm CMB}\\
    C_\ell^{\rm AGWB}\\
    C_\ell^{\rm XC}
    \end{pmatrix}\,,
\end{equation}
where $\mathbb{C}_\ell^{-1}$ is the inverse covariance matrix that here we explicit
\begin{equation}
    \mathbb{C}_\ell = 
    \begin{pmatrix}
    {\sigma_\ell^{\rm CMB}}^2 & {\sigma_\ell^{\rm CMB - AGWB}}^2 & {\sigma_\ell^{\rm CMB - XC}}^2 \\
    {\sigma_\ell^{\rm CMB - AGWB}}^2 & {\sigma_\ell^{\rm AGWB}}^2 & {\sigma_\ell^{\rm AGWB - XC}}^2 \\
    {\sigma_\ell^{\rm CMB - XC}}^2 &{\sigma_\ell^{\rm AGWB - XC}}^2 & {\sigma_\ell^{\rm XC}}^2 \\
    \end{pmatrix}\,.
\end{equation}
Here the $\sigma_\ell^2$s are the covariances and contain both the cosmic variance and the noise contribution. We report as an example
\begin{align}
        {\sigma_\ell^{\rm XC}}^2 = \frac{{C_\ell^{\rm XC}}^2+C_\ell^{\rm CMB}(C_\ell^{\rm AGWB}+N_\ell^{\rm AGWB})}{2\ell+1}
\end{align}
and
\begin{align}
    {\sigma_\ell^{\rm AGWB - XC}}^2 = \frac{2}{2\ell + 1} \left((C_\ell^{\rm AGWB} + N_\ell^{\rm AGWB}) C_\ell^{\rm XC}\right)\,.
\end{align}

We consider the angular power spectra for the noise of LISA and BBO (normalized with respect to the monopole) and we evaluate the resulting SNR. In Fig. \ref{Fig::SNR_ellmax} we report the cumulative SNR. We consider just the contribution coming from the auto- and cross- correlation, neglecting the CMB contribution (otherwise the SNR coming from the CMB, would be so high to cover the effects of the SGWB we are interested in). For both the detectors we see that only the multipoles up to $\ell \sim 4-5$ give a relevant contribution. In particular we see that LISA will not be able to detect such a signal, resulting in a SNR lower than $10^{-1}$, suggesting that the Fisher forecast will be unable to put tight constraints on $f_{\rm NL}$. On the other hand we see that BBO is able not only to reach a SNR of order one, but is so sensitive that the SNR is limited for the first multipoles just by the cosmic variance; it then saturates at a value of almost 4. This clarifies the very strong constraining power for $f_{\rm NL}$ showed in the main text but at the same time underlines how promising is the GW search and the need to improve angular resolution. The plots also show the impact of nG on the SNR. We see that effect of nG are more relevant for LISA, causing and increase (positive $f_{\rm NL}$) or a decrease (negative $f_{\rm NL}$) of the SNR. We clarify that the suppression just explained happens only for low values of $f_{\rm NL}$. In the case of a negative $f_{\rm NL}$ but with a large absolute values, assuming ($b-p$) positive, the second term of \eqref{Eq::General_NG_Bias} would give a positive dominating contribution in the auto-correlation spectrum, leading to an increase of the SNR. For the models adopted in this work, this effect would be important at values of $f_{\rm NL}$ around 10, so outside the constraints given by Planck.

\section{\label{App::AstroKernel} Astrophysical Kernel}

As reported in Subsection \ref{SubSec::GWB_Angular}, all the astrophysical information is included in the function $\mathcal{W}^{[\rm i]}[\eta(z)]$, \emph{i.e.} the astrophysical kernel. It acts as a weight function in redshift and is defined as
\begin{align}
\label{Eq::AstroKernel1}
    \mathcal{W}^{[\rm i]} [\eta(z)]\equiv \frac{f_{\rm o}}{\rho_{\rm c}} \frac{4\pi}{\overline{\Omega}_{\rm AGWB}^{[\rm i]}} \frac{w(z,\vec{\theta})N^{[\rm i]}(z,f_{\rm e},\vec{\theta})}{(1+z)}\,.
\end{align}
Here $f_{\rm o}$ is the observed frequency, $f_{\rm e}$ is the frequency at the emission and $\vec{\theta}$ encloses all the astrophysical parameter dependencies, like the masses of the two BHs, $m_1$ and $m_2$, and the merger rate today, $R_0^{\rm BBH}$, expected to be between 17.9 and 44 $\text{Gpc}^{-3} \text{yr}^{-1}$ \cite{2021arXiv211103634T}. As reported by \cite{KAGRA:2021kbb}, the upper bound on the monopole coming from BBH is $\Omega^{\rm BBH}_{\rm GW}(25\text{Hz}) = 5.0^{+1.7}_{-1.4}\times 10^{-10}$, while for the total contribution it results $3.4\times 10^{-9}$ at 25 Hz.
$N^{[\rm i]}(z,f_{\rm e},\vec{\theta})$ is the total comoving density of GWs, written as
\begin{align}
    N^{[\rm i]}(z,f_{\rm e},\vec{\theta}) = R^{[\rm i]}(z,\vec{\theta}) \frac{d E_{\rm GW,e}(f_{\rm o},z,\vec{\theta})}{df_{\rm e}d\Omega_{\rm e}}\,,
\end{align}
where $R^{[\rm i]}$ is the intrinsic comoving merger rate of BBHs and $dE_{\rm GW}/df_{\rm e}$ the energy of the GWs emitted with a frequency $f_{\rm e}$. We take into account all the stages of evolution of a binary system, \emph{i.e.} the inspiral, the merger and the ringdown, following \cite{Ajith:2007kx,Ajith:2009bn,Ajith:2012mn}. We computed also the merger rate of BBHs as follows \cite{2022JCAP...06..030B}:
\begin{itemize}
    \item We start considering the star formation rate following the phenomenological expression from \cite{Madau:2014bja}
    \begin{align}
        R_\star(z) \propto \frac{(1+z)^{\lambda_1}}{1+(\frac{1+z}{\lambda_2})^{\lambda_1+\lambda_2}}\,,
    \end{align}
    with $\lambda_1 = 2.7$ and $\lambda_2 = 2.9$ and that holds for $0\leq z \leq 8$.
    \item We then assume the binary formation rate to be proportional to the star formation rate as
    \begin{align}
        R_{\rm bin}(z) = \mathcal{C}_1 R_\star(z)
    \end{align}
    with $\mathcal{C}_1$ in the range 0 - 1 since of course not all the stars end up forming a binary.
    \item Finally we account for the time delay between the formation and the merger of the binary, so that the merger rate can be obtained after integrating over the time delay, where $p(t_{\rm d}) \propto t_{\rm d}^{-1}$ \cite{KAGRA:2021kbb}, assuming $t_{\rm d,min} = 50$ Myr for BH binaries. In order to account for the multiplicative constant, we set that the final merger rate has to be consistent with $R_{\rm BBH}$ today provided by the latest constraint from LIGO-Virgo-Kagra collaboration \cite{2021arXiv211103634T}. 
\end{itemize}
To model the BH mass distribution we consider the \textsc{Powerlaw + Peak} mass model \cite{2021arXiv211103634T}. \text

The $w(z,\vec{\theta})$ function has the role to discriminate resolved from unresolved sources in the computation, so actually it represents the efficiency of the detectors in the measurement of the AGWB. We obtain it by integrating the pdf of the SNR from 0 to the threshold that determines a detection of a resolved source, that we fix to 8 \cite{Karnesis:2021tsh} so that
\begin{align}
    w(z,\vec{\theta}) = \int_0^8 d\,{\rm SNR}\, p({\rm SNR}|z,f_{\rm o},\vec{\theta})\,.
\end{align}
We conclude that a better modeling of the window function would include the generation of a catalog of sources, also to better understand their clustering properties and better discriminating when a source can be considered resolved or not. We leave this for a more dedicated future analysis.

\newpage

\bibliographystyle{JHEP}
\bibliography{Cross_AGWB_CMB}

\providecommand{\noopsort}[1]{}\providecommand{\singleletter}[1]{#1}%

\providecommand{\href}[2]{#2}\begingroup\raggedright\begin{thebibliography}{10}

\bibitem{Sathyaprakash:2011bh}
B.~Sathyaprakash et~al., \emph{{Scientific Potential of Einstein Telescope}},
  in \emph{{46th Rencontres de Moriond on Gravitational Waves and Experimental
  Gravity}}, pp.~127--136, 8, 2011
  [\href{https://arxiv.org/abs/1108.1423}{{\ttfamily 1108.1423}}].

\bibitem{LISA:2017pwj}
{\scshape LISA} collaboration, \emph{{Laser Interferometer Space Antenna}},
  \href{https://arxiv.org/abs/1702.00786}{{\ttfamily 1702.00786}}.

\bibitem{LIGOScientific:2016wof}
{\scshape LIGO Scientific} collaboration, \emph{{Exploring the Sensitivity of
  Next Generation Gravitational Wave Detectors}},
  \href{https://doi.org/10.1088/1361-6382/aa51f4}{\emph{Class. Quant. Grav.}
  {\bfseries 34} (2017) 044001}
  [\href{https://arxiv.org/abs/1607.08697}{{\ttfamily 1607.08697}}].

\bibitem{Maggiore:2019uih}
M.~Maggiore et~al., \emph{{Science Case for the Einstein Telescope}},
  \href{https://doi.org/10.1088/1475-7516/2020/03/050}{\emph{JCAP} {\bfseries
  03} (2020) 050} [\href{https://arxiv.org/abs/1912.02622}{{\ttfamily
  1912.02622}}].

\bibitem{KAGRA:2021kbb}
{\scshape KAGRA, Virgo, LIGO Scientific} collaboration, \emph{{Upper limits on
  the isotropic gravitational-wave background from Advanced LIGO and Advanced
  Virgo\textquoteright{}s third observing run}},
  \href{https://doi.org/10.1103/PhysRevD.104.022004}{\emph{Phys. Rev. D}
  {\bfseries 104} (2021) 022004}
  [\href{https://arxiv.org/abs/2101.12130}{{\ttfamily 2101.12130}}].

\bibitem{KAGRA:2021mth}
{\scshape KAGRA, Virgo, LIGO Scientific} collaboration, \emph{{Search for
  anisotropic gravitational-wave backgrounds using data from Advanced LIGO and
  Advanced Virgo\textquoteright{}s first three observing runs}},
  \href{https://doi.org/10.1103/PhysRevD.104.022005}{\emph{Phys. Rev. D}
  {\bfseries 104} (2021) 022005}
  [\href{https://arxiv.org/abs/2103.08520}{{\ttfamily 2103.08520}}].

\bibitem{NANOGrav:2020bcs}
{\scshape NANOGrav} collaboration, \emph{{The NANOGrav 12.5 yr Data Set: Search
  for an Isotropic Stochastic Gravitational-wave Background}},
  \href{https://doi.org/10.3847/2041-8213/abd401}{\emph{Astrophys. J. Lett.}
  {\bfseries 905} (2020) L34}
  [\href{https://arxiv.org/abs/2009.04496}{{\ttfamily 2009.04496}}].

\bibitem{Ferrari:1998jf}
V.~Ferrari, S.~Matarrese and R.~Schneider, \emph{{Stochastic background of
  gravitational waves generated by a cosmological population of young, rapidly
  rotating neutron stars}},
  \href{https://doi.org/10.1046/j.1365-8711.1999.02207.x}{\emph{Mon. Not. Roy.
  Astron. Soc.} {\bfseries 303} (1999) 258}
  [\href{https://arxiv.org/abs/astro-ph/9806357}{{\ttfamily
  astro-ph/9806357}}].

\bibitem{Ferrari:1998ut}
V.~Ferrari, S.~Matarrese and R.~Schneider, \emph{{Gravitational wave background
  from a cosmological population of core collapse supernovae}},
  \href{https://doi.org/10.1046/j.1365-8711.1999.02194.x}{\emph{Mon. Not. Roy.
  Astron. Soc.} {\bfseries 303} (1999) 247}
  [\href{https://arxiv.org/abs/astro-ph/9804259}{{\ttfamily
  astro-ph/9804259}}].

\bibitem{Ignatiev:2001jr}
V.B.~Ignatiev, A.G.~Kuranov, K.A.~Postnov and M.E.~Prokhorov,
  \emph{{Gravitational wave background from coalescing compact stars in
  eccentric orbits}},
  \href{https://doi.org/10.1046/j.1365-8711.2001.04732.x}{\emph{Mon. Not. Roy.
  Astron. Soc.} {\bfseries 327} (2001) 531}
  [\href{https://arxiv.org/abs/astro-ph/0106299}{{\ttfamily
  astro-ph/0106299}}].

\bibitem{2004MNRAS.351.1237H}
E.~{Howell}, D.~{Coward}, R.~{Burman}, D.~{Blair} and J.~{Gilmore}, \emph{{The
  gravitational wave background from neutron star birth throughout the
  cosmos}},
  \href{https://doi.org/10.1111/j.1365-2966.2004.07863.x}{\emph{Monthly Notices
  of the Royal Astronomical Society} {\bfseries 351} (2004) 1237}.

\bibitem{Regimbau:2007ed}
T.~Regimbau and B.~Chauvineau, \emph{{Stochastic background from extra-galactic
  double neutron stars}},
  \href{https://doi.org/10.1088/0264-9381/24/19/S25}{\emph{Class. Quant. Grav.}
  {\bfseries 24} (2007) S627}
  [\href{https://arxiv.org/abs/0707.4327}{{\ttfamily 0707.4327}}].

\bibitem{Regimbau:2011rp}
T.~Regimbau, \emph{{The astrophysical gravitational wave stochastic
  background}}, \href{https://doi.org/10.1088/1674-4527/11/4/001}{\emph{Res.
  Astron. Astrophys.} {\bfseries 11} (2011) 369}
  [\href{https://arxiv.org/abs/1101.2762}{{\ttfamily 1101.2762}}].

\bibitem{Bartolo:2016ami}
N.~Bartolo et~al., \emph{{Science with the space-based interferometer LISA. IV:
  Probing inflation with gravitational waves}},
  \href{https://doi.org/10.1088/1475-7516/2016/12/026}{\emph{JCAP} {\bfseries
  12} (2016) 026} [\href{https://arxiv.org/abs/1610.06481}{{\ttfamily
  1610.06481}}].

\bibitem{Guzzetti:2016mkm}
M.C.~Guzzetti, N.~Bartolo, M.~Liguori and S.~Matarrese, \emph{{Gravitational
  waves from inflation}},
  \href{https://doi.org/10.1393/ncr/i2016-10127-1}{\emph{Riv. Nuovo Cim.}
  {\bfseries 39} (2016) 399}
  [\href{https://arxiv.org/abs/1605.01615}{{\ttfamily 1605.01615}}].

\bibitem{Caprini:2018mtu}
C.~Caprini and D.G.~Figueroa, \emph{{Cosmological Backgrounds of Gravitational
  Waves}}, \href{https://doi.org/10.1088/1361-6382/aac608}{\emph{Class. Quant.
  Grav.} {\bfseries 35} (2018) 163001}
  [\href{https://arxiv.org/abs/1801.04268}{{\ttfamily 1801.04268}}].

\bibitem{LISACosmologyWorkingGroup:2022jok}
{\scshape LISA Cosmology Working Group} collaboration, \emph{{Cosmology with
  the Laser Interferometer Space Antenna}},
  \href{https://arxiv.org/abs/2204.05434}{{\ttfamily 2204.05434}}.

\bibitem{Bartolo:2004if}
N.~Bartolo, E.~Komatsu, S.~Matarrese and A.~Riotto, \emph{{Non-Gaussianity from
  inflation: Theory and observations}},
  \href{https://doi.org/10.1016/j.physrep.2004.08.022}{\emph{Phys. Rept.}
  {\bfseries 402} (2004) 103}
  [\href{https://arxiv.org/abs/astro-ph/0406398}{{\ttfamily
  astro-ph/0406398}}].

\bibitem{Salopek:1990jq}
D.S.~Salopek and J.R.~Bond, \emph{{Nonlinear evolution of long wavelength
  metric fluctuations in inflationary models}},
  \href{https://doi.org/10.1103/PhysRevD.42.3936}{\emph{Phys. Rev. D}
  {\bfseries 42} (1990) 3936}.

\bibitem{Gangui:1993tt}
A.~Gangui, F.~Lucchin, S.~Matarrese and S.~Mollerach, \emph{{The Three point
  correlation function of the cosmic microwave background in inflationary
  models}}, \href{https://doi.org/10.1086/174421}{\emph{Astrophys. J.}
  {\bfseries 430} (1994) 447}
  [\href{https://arxiv.org/abs/astro-ph/9312033}{{\ttfamily
  astro-ph/9312033}}].

\bibitem{PhysRevD.63.063002}
E.~Komatsu and D.N.~Spergel, \emph{Acoustic signatures in the primary microwave
  background bispectrum},
  \href{https://doi.org/10.1103/PhysRevD.63.063002}{\emph{Phys. Rev. D}
  {\bfseries 63} (2001) 063002}.

\bibitem{Caprini:2019egz}
C.~Caprini et~al., \emph{{Detecting gravitational waves from cosmological phase
  transitions with LISA: an update}},
  \href{https://doi.org/10.1088/1475-7516/2020/03/024}{\emph{JCAP} {\bfseries
  03} (2020) 024} [\href{https://arxiv.org/abs/1910.13125}{{\ttfamily
  1910.13125}}].

\bibitem{Bartolo:2019oiq}
N.~Bartolo, D.~Bertacca, S.~Matarrese, M.~Peloso, A.~Ricciardone, A.~Riotto
  et~al., \emph{{Anisotropies and non-Gaussianity of the Cosmological
  Gravitational Wave Background}},
  \href{https://doi.org/10.1103/PhysRevD.100.121501}{\emph{Phys. Rev. D}
  {\bfseries 100} (2019) 121501}
  [\href{https://arxiv.org/abs/1908.00527}{{\ttfamily 1908.00527}}].

\bibitem{Bartolo:2019yeu}
N.~Bartolo, D.~Bertacca, S.~Matarrese, M.~Peloso, A.~Ricciardone, A.~Riotto
  et~al., \emph{{Characterizing the cosmological gravitational wave background:
  Anisotropies and non-Gaussianity}},
  \href{https://doi.org/10.1103/PhysRevD.102.023527}{\emph{Phys. Rev. D}
  {\bfseries 102} (2020) 023527}
  [\href{https://arxiv.org/abs/1912.09433}{{\ttfamily 1912.09433}}].

\bibitem{Bertacca:2019fnt}
D.~Bertacca, A.~Ricciardone, N.~Bellomo, A.C.~Jenkins, S.~Matarrese,
  A.~Raccanelli et~al., \emph{{Projection effects on the observed angular
  spectrum of the astrophysical stochastic gravitational wave background}},
  \href{https://doi.org/10.1103/PhysRevD.101.103513}{\emph{Phys. Rev. D}
  {\bfseries 101} (2020) 103513}
  [\href{https://arxiv.org/abs/1909.11627}{{\ttfamily 1909.11627}}].

\bibitem{Contaldi:2016koz}
C.R.~Contaldi, \emph{{Anisotropies of Gravitational Wave Backgrounds: A Line Of
  Sight Approach}},
  \href{https://doi.org/10.1016/j.physletb.2017.05.020}{\emph{Phys. Lett. B}
  {\bfseries 771} (2017) 9} [\href{https://arxiv.org/abs/1609.08168}{{\ttfamily
  1609.08168}}].

\bibitem{Cusin:2019jhg}
G.~Cusin, I.~Dvorkin, C.~Pitrou and J.-P.~Uzan, \emph{{Stochastic gravitational
  wave background anisotropies in the mHz band: astrophysical dependencies}},
  \href{https://doi.org/10.1093/mnrasl/slz182}{\emph{Mon. Not. Roy. Astron.
  Soc.} {\bfseries 493} (2020) L1}
  [\href{https://arxiv.org/abs/1904.07757}{{\ttfamily 1904.07757}}].

\bibitem{dodelson2020modern}
S.~Dodelson and F.~Schmidt, \emph{Modern cosmology}, Academic Press (2020).

\bibitem{Bartolo:2006cu}
N.~Bartolo, S.~Matarrese and A.~Riotto, \emph{{CMB Anisotropies at Second Order
  I}}, \href{https://doi.org/10.1088/1475-7516/2006/06/024}{\emph{JCAP}
  {\bfseries 06} (2006) 024}
  [\href{https://arxiv.org/abs/astro-ph/0604416}{{\ttfamily
  astro-ph/0604416}}].

\bibitem{Bartolo:2006fj}
N.~Bartolo, S.~Matarrese and A.~Riotto, \emph{{CMB Anisotropies at
  Second-Order. 2. Analytical Approach}},
  \href{https://doi.org/10.1088/1475-7516/2007/01/019}{\emph{JCAP} {\bfseries
  01} (2007) 019} [\href{https://arxiv.org/abs/astro-ph/0610110}{{\ttfamily
  astro-ph/0610110}}].

\bibitem{Ricciardone:2021kel}
A.~Ricciardone, L.V.~Dall'Armi, N.~Bartolo, D.~Bertacca, M.~Liguori and
  S.~Matarrese, \emph{{Cross-Correlating Astrophysical and Cosmological
  Gravitational Wave Backgrounds with the Cosmic Microwave Background}},
  \href{https://doi.org/10.1103/PhysRevLett.127.271301}{\emph{Phys. Rev. Lett.}
  {\bfseries 127} (2021) 271301}
  [\href{https://arxiv.org/abs/2106.02591}{{\ttfamily 2106.02591}}].

\bibitem{1967ApJ...147...73S}
R.K.~{Sachs} and A.M.~{Wolfe}, \emph{{Perturbations of a Cosmological Model and
  Angular Variations of the Microwave Background}},
  \href{https://doi.org/10.1086/148982}{\emph{The Astrophysical Journal}
  {\bfseries 147} (1967) 73}.

\bibitem{Regimbau:2016ike}
T.~Regimbau, M.~Evans, N.~Christensen, E.~Katsavounidis, B.~Sathyaprakash and
  S.~Vitale, \emph{{Digging deeper: Observing primordial gravitational waves
  below the binary black hole produced stochastic background}},
  \href{https://doi.org/10.1103/PhysRevLett.118.151105}{\emph{Phys. Rev. Lett.}
  {\bfseries 118} (2017) 151105}
  [\href{https://arxiv.org/abs/1611.08943}{{\ttfamily 1611.08943}}].

\bibitem{Bardeen:1985tr}
J.M.~Bardeen, J.R.~Bond, N.~Kaiser and A.S.~Szalay, \emph{{The Statistics of
  Peaks of Gaussian Random Fields}},
  \href{https://doi.org/10.1086/164143}{\emph{Astrophys. J.} {\bfseries 304}
  (1986) 15}.

\bibitem{Dalal:2007cu}
N.~Dalal, O.~Dore, D.~Huterer and A.~Shirokov, \emph{{The imprints of
  primordial non-gaussianities on large-scale structure: scale dependent bias
  and abundance of virialized objects}},
  \href{https://doi.org/10.1103/PhysRevD.77.123514}{\emph{Phys. Rev. D}
  {\bfseries 77} (2008) 123514}
  [\href{https://arxiv.org/abs/0710.4560}{{\ttfamily 0710.4560}}].

\bibitem{Matarrese:2008nc}
S.~Matarrese and L.~Verde, \emph{{The effect of primordial non-Gaussianity on
  halo bias}}, \href{https://doi.org/10.1086/587840}{\emph{Astrophys. J. Lett.}
  {\bfseries 677} (2008) L77}
  [\href{https://arxiv.org/abs/0801.4826}{{\ttfamily 0801.4826}}].

\bibitem{Slosar:2008hx}
A.~Slosar, C.~Hirata, U.~Seljak, S.~Ho and N.~Padmanabhan, \emph{{Constraints
  on local primordial non-Gaussianity from large scale structure}},
  \href{https://doi.org/10.1088/1475-7516/2008/08/031}{\emph{JCAP} {\bfseries
  08} (2008) 031} [\href{https://arxiv.org/abs/0805.3580}{{\ttfamily
  0805.3580}}].

\bibitem{Planck:2019kim}
{\scshape Planck} collaboration, \emph{{Planck 2018 results. IX. Constraints on
  primordial non-Gaussianity}},
  \href{https://doi.org/10.1051/0004-6361/201935891}{\emph{Astron. Astrophys.}
  {\bfseries 641} (2020) A9}
  [\href{https://arxiv.org/abs/1905.05697}{{\ttfamily 1905.05697}}].

\bibitem{Tucci:2016hng}
M.~Tucci, V.~Desjacques and M.~Kunz, \emph{{Cosmic Infrared Background
  anisotropies as a window into primordial non-Gaussianity}},
  \href{https://doi.org/10.1093/mnras/stw2086}{\emph{Mon. Not. Roy. Astron.
  Soc.} {\bfseries 463} (2016) 2046}
  [\href{https://arxiv.org/abs/1606.02323}{{\ttfamily 1606.02323}}].

\bibitem{Giannantonio:2013uqa}
T.~Giannantonio, A.J.~Ross, W.J.~Percival, R.~Crittenden, D.~Bacher,
  M.~Kilbinger et~al., \emph{{Improved Primordial Non-Gaussianity Constraints
  from Measurements of Galaxy Clustering and the Integrated Sachs-Wolfe
  Effect}}, \href{https://doi.org/10.1103/PhysRevD.89.023511}{\emph{Phys. Rev.
  D} {\bfseries 89} (2014) 023511}
  [\href{https://arxiv.org/abs/1303.1349}{{\ttfamily 1303.1349}}].

\bibitem{Bellomo:2020pnw}
N.~Bellomo, J.L.~Bernal, G.~Scelfo, A.~Raccanelli and L.~Verde, \emph{{Beware
  of commonly used approximations. Part I. Errors in forecasts}},
  \href{https://doi.org/10.1088/1475-7516/2020/10/016}{\emph{JCAP} {\bfseries
  10} (2020) 016} [\href{https://arxiv.org/abs/2005.10384}{{\ttfamily
  2005.10384}}].

\bibitem{2011arXiv1104.2932L}
J.~{Lesgourgues}, \emph{{The Cosmic Linear Anisotropy Solving System (CLASS) I:
  Overview}}, \href{https://doi.org/10.48550/arXiv.1104.2932}{\emph{arXiv
  e-prints} (2011) arXiv:1104.2932}
  [\href{https://arxiv.org/abs/1104.2932}{{\ttfamily 1104.2932}}].

\bibitem{2021arXiv211103634T}
{The LIGO Scientific Collaboration}, {the Virgo Collaboration} and {the KAGRA
  Collaboration}, \emph{{The population of merging compact binaries inferred
  using gravitational waves through GWTC-3}},
  \href{https://doi.org/10.48550/arXiv.2111.03634}{\emph{arXiv e-prints} (2021)
  arXiv:2111.03634} [\href{https://arxiv.org/abs/2111.03634}{{\ttfamily
  2111.03634}}].

\bibitem{Madau:2014bja}
P.~Madau and M.~Dickinson, \emph{{Cosmic Star Formation History}},
  \href{https://doi.org/10.1146/annurev-astro-081811-125615}{\emph{Ann. Rev.
  Astron. Astrophys.} {\bfseries 52} (2014) 415}
  [\href{https://arxiv.org/abs/1403.0007}{{\ttfamily 1403.0007}}].

\bibitem{Ajith:2007kx}
P.~Ajith et~al., \emph{{A Template bank for gravitational waveforms from
  coalescing binary black holes. I. Non-spinning binaries}},
  \href{https://doi.org/10.1103/PhysRevD.77.104017}{\emph{Phys. Rev. D}
  {\bfseries 77} (2008) 104017}
  [\href{https://arxiv.org/abs/0710.2335}{{\ttfamily 0710.2335}}].

\bibitem{Ajith:2009bn}
P.~Ajith et~al., \emph{{Inspiral-merger-ringdown waveforms for black-hole
  binaries with non-precessing spins}},
  \href{https://doi.org/10.1103/PhysRevLett.106.241101}{\emph{Phys. Rev. Lett.}
  {\bfseries 106} (2011) 241101}
  [\href{https://arxiv.org/abs/0909.2867}{{\ttfamily 0909.2867}}].

\bibitem{Ajith:2012mn}
P.~Ajith, N.~Fotopoulos, S.~Privitera, A.~Neunzert and A.J.~Weinstein,
  \emph{{Effectual template bank for the detection of gravitational waves from
  inspiralling compact binaries with generic spins}},
  \href{https://doi.org/10.1103/PhysRevD.89.084041}{\emph{Phys. Rev. D}
  {\bfseries 89} (2014) 084041}
  [\href{https://arxiv.org/abs/1210.6666}{{\ttfamily 1210.6666}}].

\bibitem{Corbin:2005ny}
V.~Corbin and N.J.~Cornish, \emph{{Detecting the cosmic gravitational wave
  background with the big bang observer}},
  \href{https://doi.org/10.1088/0264-9381/23/7/014}{\emph{Class. Quant. Grav.}
  {\bfseries 23} (2006) 2435}
  [\href{https://arxiv.org/abs/gr-qc/0512039}{{\ttfamily gr-qc/0512039}}].

\bibitem{Cusin:2017fwz}
G.~Cusin, C.~Pitrou and J.-P.~Uzan, \emph{{Anisotropy of the astrophysical
  gravitational wave background: Analytic expression of the angular power
  spectrum and correlation with cosmological observations}},
  \href{https://doi.org/10.1103/PhysRevD.96.103019}{\emph{Phys. Rev. D}
  {\bfseries 96} (2017) 103019}
  [\href{https://arxiv.org/abs/1704.06184}{{\ttfamily 1704.06184}}].

\bibitem{Jenkins:2018uac}
A.C.~Jenkins, M.~Sakellariadou, T.~Regimbau and E.~Slezak, \emph{{Anisotropies
  in the astrophysical gravitational-wave background: Predictions for the
  detection of compact binaries by LIGO and Virgo}},
  \href{https://doi.org/10.1103/PhysRevD.98.063501}{\emph{Phys. Rev. D}
  {\bfseries 98} (2018) 063501}
  [\href{https://arxiv.org/abs/1806.01718}{{\ttfamily 1806.01718}}].

\bibitem{Schmidt:2012ne}
F.~Schmidt and D.~Jeong, \emph{{Cosmic Rulers}},
  \href{https://doi.org/10.1103/PhysRevD.86.083527}{\emph{Phys. Rev. D}
  {\bfseries 86} (2012) 083527}
  [\href{https://arxiv.org/abs/1204.3625}{{\ttfamily 1204.3625}}].

\bibitem{2022JCAP...06..030B}
N.~{Bellomo}, D.~{Bertacca}, A.C.~{Jenkins}, S.~{Matarrese}, A.~{Raccanelli},
  T.~{Regimbau} et~al., \emph{{CLASS\_GWB: robust modeling of the astrophysical
  gravitational wave background anisotropies}},
  \href{https://doi.org/10.1088/1475-7516/2022/06/030}{\emph{The Astrophysical
  Journal} {\bfseries 2022} (2022) 030}
  [\href{https://arxiv.org/abs/2110.15059}{{\ttfamily 2110.15059}}].

\bibitem{2022JCAP...11..009B}
N.~{Bartolo}, D.~{Bertacca}, R.~{Caldwell}, C.R.~{Contaldi}, G.~{Cusin}, V.~{De
  Luca} et~al., \emph{{Probing anisotropies of the Stochastic Gravitational
  Wave Background with LISA}},
  \href{https://doi.org/10.1088/1475-7516/2022/11/009}{\emph{JCAP} {\bfseries
  2022} (2022) 009} [\href{https://arxiv.org/abs/2201.08782}{{\ttfamily
  2201.08782}}].

\bibitem{Jeong:2011as}
D.~Jeong, F.~Schmidt and C.M.~Hirata, \emph{{Large-scale clustering of galaxies
  in general relativity}},
  \href{https://doi.org/10.1103/PhysRevD.85.023504}{\emph{Phys. Rev. D}
  {\bfseries 85} (2012) 023504}
  [\href{https://arxiv.org/abs/1107.5427}{{\ttfamily 1107.5427}}].

\bibitem{Bertacca:2012tp}
D.~Bertacca, R.~Maartens, A.~Raccanelli and C.~Clarkson, \emph{{Beyond the
  plane-parallel and Newtonian approach: Wide-angle redshift distortions and
  convergence in general relativity}},
  \href{https://doi.org/10.1088/1475-7516/2012/10/025}{\emph{JCAP} {\bfseries
  10} (2012) 025} [\href{https://arxiv.org/abs/1205.5221}{{\ttfamily
  1205.5221}}].

\bibitem{Scelfo:2018sny}
G.~Scelfo, N.~Bellomo, A.~Raccanelli, S.~Matarrese and L.~Verde,
  \emph{{GW$\times$LSS: chasing the progenitors of merging binary black
  holes}}, \href{https://doi.org/10.1088/1475-7516/2018/09/039}{\emph{JCAP}
  {\bfseries 09} (2018) 039}
  [\href{https://arxiv.org/abs/1809.03528}{{\ttfamily 1809.03528}}].

\bibitem{Mukherjee:2019wcg}
S.~Mukherjee, B.D.~Wandelt and J.~Silk, \emph{{Probing the theory of gravity
  with gravitational lensing of gravitational waves and galaxy surveys}},
  \href{https://doi.org/10.1093/mnras/staa827}{\emph{Mon. Not. Roy. Astron.
  Soc.} {\bfseries 494} (2020) 1956}
  [\href{https://arxiv.org/abs/1908.08951}{{\ttfamily 1908.08951}}].

\bibitem{Libanore:2020fim}
S.~Libanore, M.C.~Artale, D.~Karagiannis, M.~Liguori, N.~Bartolo, Y.~Bouffanais
  et~al., \emph{{Gravitational Wave mergers as tracers of Large Scale
  Structures}},
  \href{https://doi.org/10.1088/1475-7516/2021/02/035}{\emph{JCAP} {\bfseries
  02} (2021) 035} [\href{https://arxiv.org/abs/2007.06905}{{\ttfamily
  2007.06905}}].

\bibitem{Planck:2018vyg}
{\scshape Planck} collaboration, \emph{{Planck 2018 results. VI. Cosmological
  parameters}},
  \href{https://doi.org/10.1051/0004-6361/201833910}{\emph{Astron. Astrophys.}
  {\bfseries 641} (2020) A6}
  [\href{https://arxiv.org/abs/1807.06209}{{\ttfamily 1807.06209}}].

\bibitem{Penzias:1965wn}
A.A.~Penzias and R.W.~Wilson, \emph{{A Measurement of excess antenna
  temperature at 4080-Mc/s}},
  \href{https://doi.org/10.1086/148307}{\emph{Astrophys. J.} {\bfseries 142}
  (1965) 419}.

\bibitem{Planck:2019nip}
{\scshape Planck} collaboration, \emph{{Planck 2018 results. V. CMB power
  spectra and likelihoods}},
  \href{https://doi.org/10.1051/0004-6361/201936386}{\emph{Astron. Astrophys.}
  {\bfseries 641} (2020) A5}
  [\href{https://arxiv.org/abs/1907.12875}{{\ttfamily 1907.12875}}].

\bibitem{1995PhDT..........H}
W.T.~{Hu}, \emph{{Wandering in the Background: a Cosmic Microwave Background
  Explorer}}, Ph.D. thesis, University of California, Berkeley, Jan., 1995.

\bibitem{Kaiser:1984sw}
N.~Kaiser, \emph{{On the Spatial correlations of Abell clusters}},
  \href{https://doi.org/10.1086/184341}{\emph{Astrophys. J. Lett.} {\bfseries
  284} (1984) L9}.

\bibitem{Efstathiou:1988tk}
G.~Efstathiou, C.S.~Frenk, S.D.M.~White and M.~Davis, \emph{{Gravitational
  clustering from scale free initial conditions}}, {\emph{Mon. Not. Roy.
  Astron. Soc.} {\bfseries 235} (1988) 715}.

\bibitem{Barreira:2021ueb}
A.~Barreira, \emph{{Predictions for local PNG bias in the galaxy power spectrum
  and bispectrum and the consequences for f $_{NL}$ constraints}},
  \href{https://doi.org/10.1088/1475-7516/2022/01/033}{\emph{JCAP} {\bfseries
  01} (2022) 033} [\href{https://arxiv.org/abs/2107.06887}{{\ttfamily
  2107.06887}}].

\bibitem{Lazeyras:2022koc}
T.~Lazeyras, A.~Barreira, F.~Schmidt and V.~Desjacques, \emph{{Assembly bias in
  the local PNG halo bias and its implication for f $_{NL}$ constraints}},
  \href{https://doi.org/10.1088/1475-7516/2023/01/023}{\emph{JCAP} {\bfseries
  01} (2023) 023} [\href{https://arxiv.org/abs/2209.07251}{{\ttfamily
  2209.07251}}].

\bibitem{Barreira:2022sey}
A.~Barreira, \emph{{Can we actually constrain f$_{NL}$ using the
  scale-dependent bias effect? An illustration of the impact of galaxy bias
  uncertainties using the BOSS DR12 galaxy power spectrum}},
  \href{https://doi.org/10.1088/1475-7516/2022/11/013}{\emph{JCAP} {\bfseries
  11} (2022) 013} [\href{https://arxiv.org/abs/2205.05673}{{\ttfamily
  2205.05673}}].

\bibitem{Blas:2011rf}
D.~Blas, J.~Lesgourgues and T.~Tram, \emph{{The Cosmic Linear Anisotropy
  Solving System (CLASS) II: Approximation schemes}},
  \href{https://doi.org/10.1088/1475-7516/2011/07/034}{\emph{JCAP} {\bfseries
  07} (2011) 034} [\href{https://arxiv.org/abs/1104.2933}{{\ttfamily
  1104.2933}}].

\bibitem{DiDio:2013bqa}
E.~Di~Dio, F.~Montanari, J.~Lesgourgues and R.~Durrer, \emph{{The CLASSgal code
  for Relativistic Cosmological Large Scale Structure}},
  \href{https://doi.org/10.1088/1475-7516/2013/11/044}{\emph{JCAP} {\bfseries
  11} (2013) 044} [\href{https://arxiv.org/abs/1307.1459}{{\ttfamily
  1307.1459}}].

\bibitem{Desjacques:2020zue}
V.~Desjacques, Y.B.~Ginat and R.~Reischke, \emph{{Statistics of a single sky:
  constrained random fields and the imprint of Bardeen potentials on galaxy
  clustering}},  \href{https://arxiv.org/abs/2009.02036}{{\ttfamily
  2009.02036}}.

\bibitem{Tegmark:1999ke}
M.~Tegmark, D.J.~Eisenstein, W.~Hu and A.~de~Oliveira-Costa, \emph{{Foregrounds
  and forecasts for the cosmic microwave background}},
  \href{https://doi.org/10.1086/308348}{\emph{Astrophys. J.} {\bfseries 530}
  (2000) 133} [\href{https://arxiv.org/abs/astro-ph/9905257}{{\ttfamily
  astro-ph/9905257}}].

\bibitem{Karnesis:2021tsh}
N.~Karnesis, S.~Babak, M.~Pieroni, N.~Cornish and T.~Littenberg,
  \emph{{Characterization of the stochastic signal originating from compact
  binary populations as measured by LISA}},
  \href{https://doi.org/10.1103/PhysRevD.104.043019}{\emph{Phys. Rev. D}
  {\bfseries 104} (2021) 043019}
  [\href{https://arxiv.org/abs/2103.14598}{{\ttfamily 2103.14598}}].

\end{thebibliography}\endgroup

\end{document}